\RequirePackage{fix-cm}
\documentclass[twocolumn]{svjour3} 

\smartqed

\usepackage[english]{babel} 
\usepackage[T1]{fontenc} 
\usepackage[ansinew]{inputenc} 
\usepackage{lmodern} 
\usepackage{hyperref}
\addto\extrasenglish{
    
}
\addto\extrasenglish{
    
}
\usepackage{graphicx}
\usepackage{listings}
\usepackage{xcolor}
\usepackage{colortbl}
\usepackage{subfig}
\usepackage{amsfonts}
\usepackage{amsmath}
\usepackage{amssymb}
\usepackage{booktabs} 
\usepackage{hhline}
\usepackage{etoolbox}
\usepackage{multirow}
\usepackage{float}
\usepackage{natbib}
\usepackage{hhline}

\usepackage{listings}
\definecolor{LightCyan}{rgb}{0.88,1,1}
\definecolor{mygreen}{RGB}{28,172,0} 
\definecolor{mylilas}{RGB}{170,55,241}
\journalname{}

\begin{document}

\title{A new generation 99 line Matlab code for compliance Topology Optimization and its extension to 3D}

\titlerunning{A new 99 line Matlab code for compliance Topology Optimization}

\author{Federico Ferrari \and
        Ole Sigmund}
\institute{Department of Mechanical Engineering \\
           Technical University of Denmark \\
           Nils Koppels All\'{e} 404, 2800 Kongens Lyngby, Denmark \\
           \email{fedeferro@hotmail.it, sigmund@mek.dtu.dk} \\
           }
           
\date{\textbf{Accepted paper, to appear soon in Structural and Multidisciplinary Optimization}}

\maketitle

\begin{abstract}
Compact and efficient Matlab implementations of compliance Topology Optimization (TO) for 2D and 3D continua are given, consisting of 99 and 125 lines respectively. On discretizations ranging from $3\cdot 10^{4}$ to $4.8\cdot10^{5}$ elements, the 2D version, named \texttt{top99neo}, shows speedups from 2.55 to 5.5 times compared to the well--known \texttt{top88} code \citep{andreassen-etal_11a}. The 3D version, named \texttt{top3D125}, is the most compact and efficient Matlab implementation for 3D TO to date, showing a speedup of 1.9 times compared to the code of \cite{amir-etal_14a}, on a discretization with $2.2\cdot10^{5}$ elements. For both codes, improvements are due to much more efficient procedures for the assembly and implementation of filters and shortcuts in the design update step. The use of an acceleration strategy, yielding major cuts in the overall computational time, is also discussed, stressing its easy integration within the basic codes.
\end{abstract}

\section{Introduction}
 \label{Sec:intro}

The celebrated \texttt{top99} Matlab code developed by \citet{sigmund_01a} has certainly promoted the spreading of Topology Optimization among engineers and researchers, and the speedups carried by its heir, \texttt{top88} \citep{andreassen-etal_11a}, substantially increased the scale of examples that can be solved on a laptop.

On these footprints, several other codes have followed, involving extension to 3D problems \citep{liu-tovar_14a,amir-etal_14a}, material design \citep{andreassen-andreasen_14a,xia-breitkopf_15a}, level--set parametrizations  \citep{wang_07a,challis_10a}, use of advanced discretization techniques \citep{talischi-etal_12a,suresh_10a,sanders-etal_18a}, or integration of TO within some finite element frameworks.

With the evolution of TO and its application to more and more challenging problems, implementations in \texttt{top88} may have become outdated. Also Matlab has improved in the last decade. Hence, we believe it is time to present a new ``exemplary'' code collecting shortcuts and speedups, allowing to tackle medium/large--scale TO problems efficiently on a laptop. Preconditioned iterative solvers, applied for example in \citep{amir-sigmund_11a, amir-etal_14a} and \citep{ferrari-etal_18a,ferrari-sigmund_20a} allow the solution of the state equation with nearly optimal efficiency \citep{book:saad92}. Thus, the computational bottleneck has been shifted on other operations, such as the matrix assembly or the repeated application of filters. Efficiency improvements for these operations were touched upon by \cite{andreassen-etal_11a}, however without giving a quantitative analysis about time and memory savings.

\begin{figure*}[t]
 \centering
  \subfloat[]{\fbox{
   \includegraphics[scale = 0.25, keepaspectratio]
   {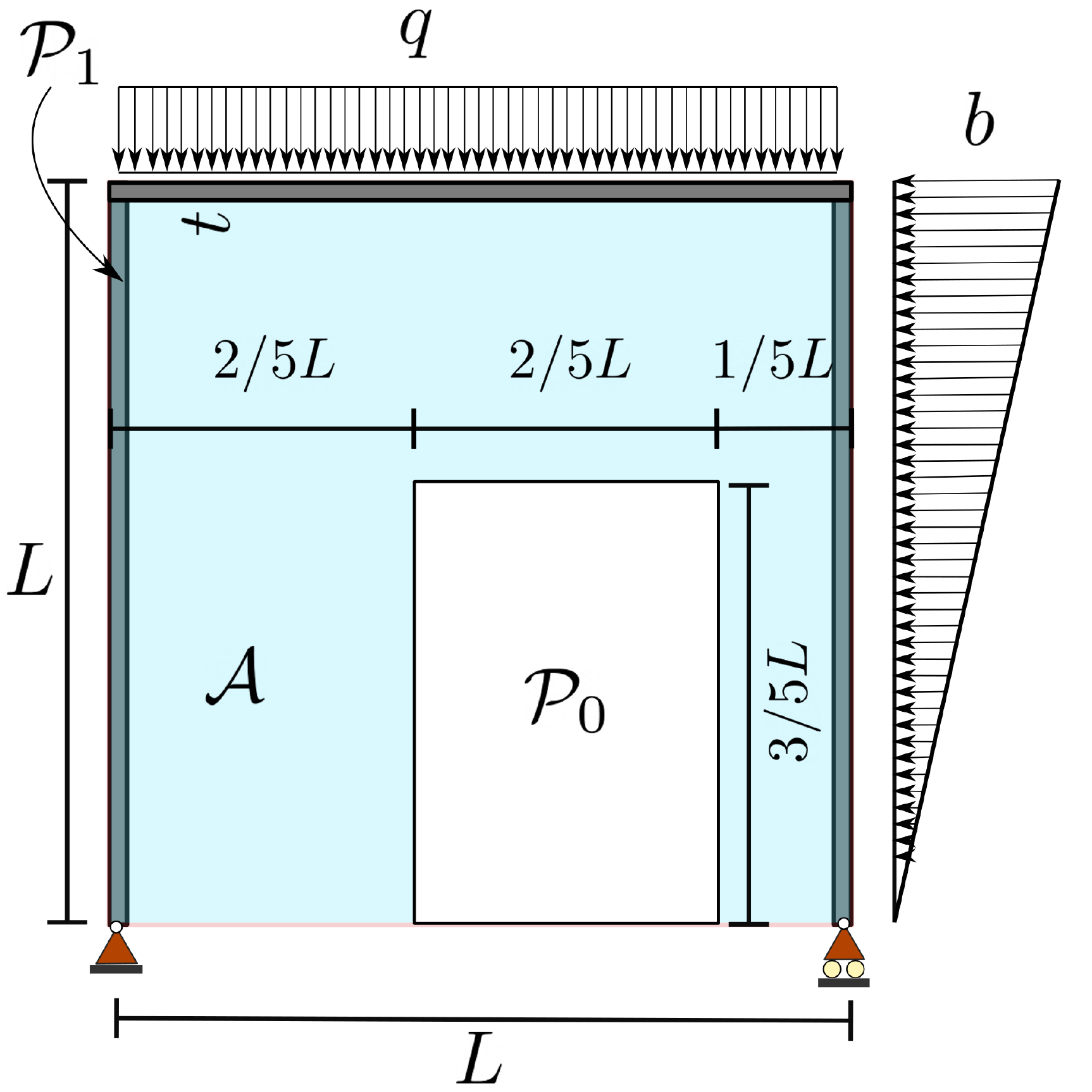}}} \qquad
  \subfloat[]{\fbox{
   \includegraphics[scale = 0.385, keepaspectratio]
   {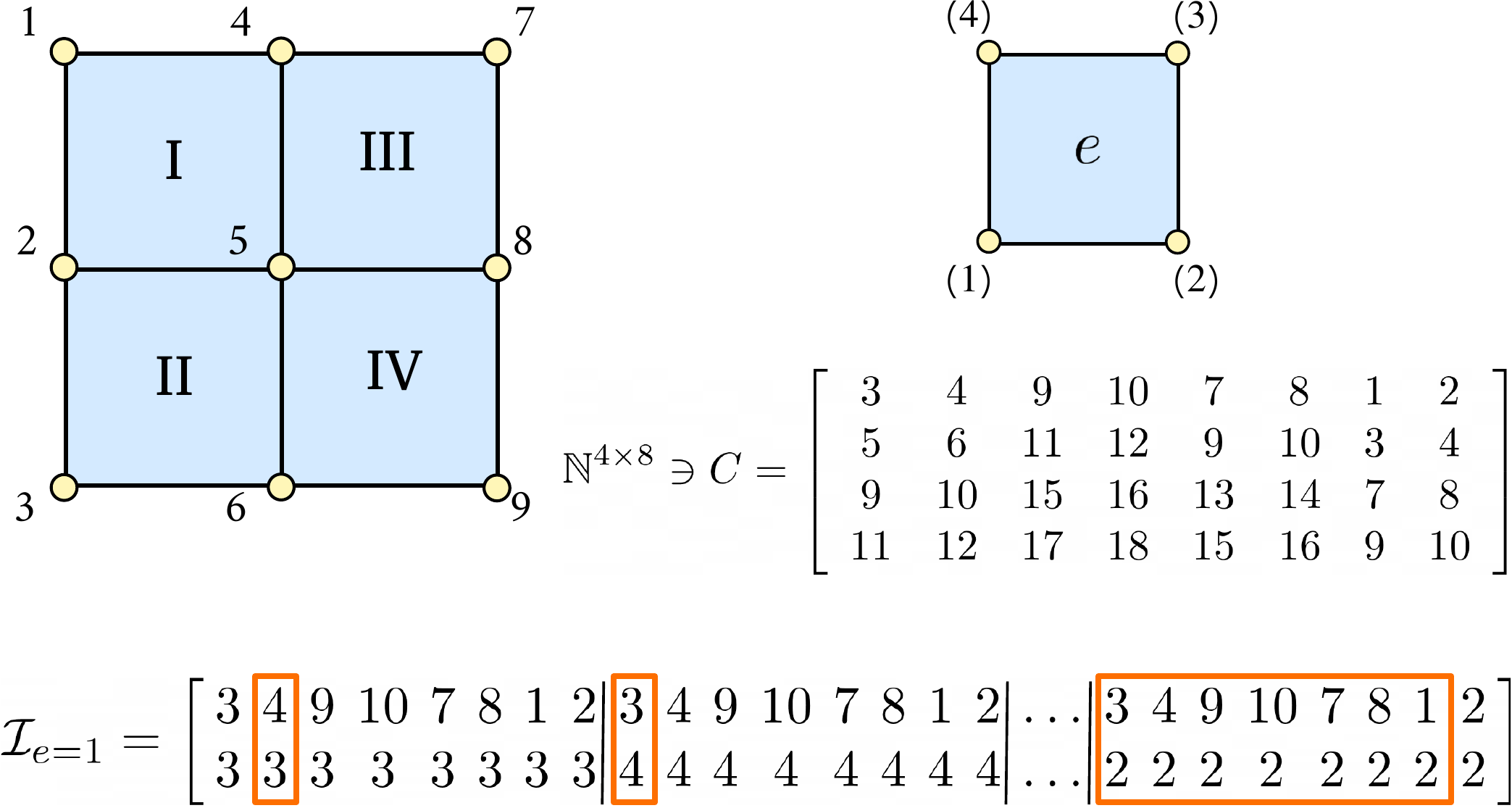}}}
 \caption{\small{Definition of the active $\mathcal{A}$, passive solid $\mathcal{P}_{1}$ and void $\mathcal{P}_{0}$ domains (a) and illustration of the connectivity matrix $C$ for a simple discretization (b). The set of indices $\mathcal{I}$, here shown for the element $e=1$, is used by the assembly operation. The symmetric repetitions in $\mathcal{I}$ are highlighted, and their elimination gives the reduced set $\mathcal{I}_{r}$ (see \autoref{sSec:speedupAssembly})}}
 \label{fig:settingFigures}
\end{figure*}

Here we provide compact Matlab codes for minimum compliance Topology Optimization of 2D and 3D continua which show a substantial speedup compared to the \texttt{top88} code. We include several extensions by default, such as specification of passive domains, a volume--preserving density projection \citep{guest-etal_04a,wang-etal_11a} and continuation strategies for the penalization and projection parameters in a very compact, yet sharp, implementation. Coincidentally, the new 2D TO implementation consist of 99 lines of code, and is thus named \texttt{top99neo}. We also show how to include an acceleration technique recently investigated for TO by \citep{thesis:li-paulino_18}, with a few extra lines of code and potentially carrying major speedups. Changes needed for the extension to 3D problems are remarkably small, making the corresponding code (\texttt{top3d125}) the most compact and efficient Matlab implementation for 3D compliance TO to date.

Our primary goal is not to present innovative new research. Rather, we aim at sharing some shortcuts and speedups that we have noticed through time, to the benefit of the research community. Improvements introduced by the present codes will be much useful also on more advanced problems, such as buckling optimization, which will be dealt with in an upcoming work.

The paper is organized as follows. In \autoref{Sec:Formulations} we recall the setting of TO for minimum compliance. \autoref{Sec:codeStructure} is devoted to describe the overall structure of the 2D code, focusing on differences with respect to \texttt{top88}. \autoref{sSec:speedupAssembly}--\autoref{sSec:FrameReinforcement} give insights about the main speedups and show performance improvements with respect to \texttt{top88}. The very few changes needed for the 3D code are listed in \autoref{Sec:3Dextension}, where an example is presented and the efficiency is compared to the previous code from \cite{amir-etal_14a}. Some final remarks are given in \autoref{Sec:conclusions}. \autoref{Sec:lmEstimate} gives some details about the redesigns step that are useful for better understanding a method proposed in \autoref{sSec:filtersVP} and the Matlab codes are listed in \autoref{Sec:2DcodeComplianceMinimization} and \autoref{Sec:3DcodeComplianceMinimization}.

\section{Problem formulation and solution scheme}
 \label{Sec:Formulations}

We consider a 2D/3D discretization $\Omega_{h}$ consisting of $m$ equi--sized quadrilateral elements $\Omega_{e}$. Hereafter we denote by $n$ the global number of Degrees of Freedom (DOFs) in the discretization and by $d$ the number of (local) DOFs of each element.

Let $\mathbf{x}= \{ x_{e} \}_{e = 1:m}\in[0, 1]^{m}$ be partitioned between $\mathbf{x}_{\mathcal{A}}$ and $\mathbf{x}_{\mathcal{P}}$, the sets of active (design) variables and passive elements, respectively. The latter may be further split in the sets of passive solid $\mathcal{P}_{1}$ ($x_{e}=1$) and void $\mathcal{P}_{0}$ ($x_{e}=0$) elements, of cardinalities $m_{\mathcal{P}_{1}}$ and $m_{\mathcal{P}_{0}}$, respectively (see \autoref{fig:settingFigures}(a)).

The set of physical variables $\hat{\mathbf{x}}_{\mathcal{A}} = \mathcal{H}(\tilde{\mathbf{x}})$ are defined by the relaxed Heaviside projection \citep{wang-etal_11a}
\begin{equation}
 \label{eq:projection}
  \mathcal{H}( \tilde{x}_{e}, \eta, \beta ) = \frac{\tanh(\beta\eta) + \tanh(\beta(\tilde{x}_{e}-\eta))}{\tanh(\beta\eta)+\tanh(\beta(1-\eta))}
\end{equation}
with threshhold $\eta$ and sharpness factor $\beta$, where $\tilde{\mathbf{x}} = H\mathbf{x}$ is the filtered field, obtained by the linear operator
\begin{equation}
 \label{eq:filter}
  H\left( x_{e}, r_{\rm min} \right) := \frac{\sum_{i \in \mathcal{N}_{e}}h_{e,i}x_{i}}
   {\sum_{i \in \mathcal{N}_{e}}h_{e,i}}
\end{equation}
where $\mathcal{N}_{e} = \{ i \mid {\rm dist}( \Omega_{i}, \Omega_{e} ) \leq r_{\rm min} \}$ and $h_{e,i} = \max( 0, r_{\rm min} - {\rm dist}( \Omega_{i}, \Omega_{e} ) )$.

Given a load vector $\mathbf{f}\in\mathbb{R}^{n}$ and the volume fraction $f \in (0, 1)$ we consider the optimization problem
\begin{equation}
 \label{eq:OptimizationProblem}
  \begin{cases}
              & \min\limits_{\mathbf{x}_{\mathcal{A}}\in\left[ 0, 1 \right]^{m_{\mathcal{A}}}} 
              c\left( \hat{\mathbf{x}} \right) \\
   {\rm s.t.} & V\left( \hat{\mathbf{x}} \right) \leq f|\Omega_{h}|
  \end{cases}
\end{equation}
for the minimization of compliance $c\left(\hat{\mathbf{x}}\right) = \mathbf{u}^{T}\mathbf{f}$ with an upper bound on the overall volume
\begin{equation}
 \label{eq:StructuralVolume}
  V\left( \hat{\mathbf{x}} \right) = \sum^{m}_{e=1} |\Omega_{e}| \hat{x}_{e} = \frac{1}{m}
  \left( m_{\mathcal{P}_{1}} + \sum_{e\in\mathcal{A}} \hat{x}_{e} \right) \leq f
\end{equation}

Problem \eqref{eq:OptimizationProblem} is solved with a nested iterative loop. At each iteration, the displacement $\mathbf{u}$ is computed by solving the equilibrium problem
\begin{equation}
 \label{eq:stateProblem}
  K\mathbf{u} = \mathbf{f}
\end{equation}
where the stiffness matrix $K = K(\hat{\mathbf{x}})$ depends on the physical variables through a SIMP interpolation \citep{bendsoe-sigmund_99a} of the Young modulus
\begin{equation}
 \label{eq:simpPenalization}
  E(\hat{x}_{e}) = E_{\rm min} + \hat{x}^{p}_{e}( E_{0} - E_{\rm min})
\end{equation}
with $E_{0}$ and $E_{\rm min}$ the moduli of solid and void ($E_{\rm min} \ll E_{0}$). The gradients of compliance and structural volume with respect to $\hat{\mathbf{x}}$ read ($\chi_{e} = 1$ if $e\in\mathcal{A}$ and 0 otherwise and $\mathbf{1}_{m}$ is the identity vector of dimension $m$)
\begin{equation}
 \label{eq:sensitivityComplianceVolume}
  \nabla_{\mathbf{\hat{x}}} c\left( \hat{\mathbf{x}} \right) =
  - \mathbf{u}^{T}\nabla_{\mathbf{\hat{x}}}K\mathbf{u}\chi_{\mathcal{A}} \: , \quad
  \nabla_{\hat{\mathbf{x}}} V\left(\hat{\mathbf{x}}\right) = \frac{1}{m}\mathbf{1}_{m}\chi_{\mathcal{A}}
\end{equation}
and the sensitivities with respect to the design variables are recovered as
\begin{equation}
 \label{eq:sensitivityComplianceVolumeDV}
  \begin{aligned}
   \nabla_{\mathbf{x}} c\left( \mathbf{x} \right) & =
   \nabla_{\tilde{\mathbf{x}}}\mathcal{H} \odot
   (H^{T}\nabla_{\mathbf{\hat{x}}} c\left( \hat{\mathbf{x}} \right))
   \\
   \nabla_{\mathbf{x}} V\left( \mathbf{x} \right) & =
   \nabla_{\tilde{\mathbf{x}}}\mathcal{H} \odot
   (H^{T}\nabla_{\mathbf{\hat{x}}} V\left( \hat{\mathbf{x}} \right))
  \end{aligned}
\end{equation}
where $\odot$ represents the elementwise multiplication and
\begin{equation}
 \nabla_{\tilde{\mathbf{x}}}\mathcal{H} = \beta\frac{1-\tanh(\beta(\tilde{\mathbf{x}}-\eta))^{2}}
 {\tanh(\beta\eta)+\tanh(\beta(1-\eta))}
\end{equation}

The active design variables $e\in\mathcal{A}$ are then updated by the Optimality Criterion rule \citep{sigmund_01a}
\begin{equation}
 \label{eq:OCupdate}
 x_{k+1,e} = \mathcal{U}(x_{k, e}) =
 \begin{cases}
  \delta_{-} & {\rm if} \: \mathcal{F}_{k,e} < \delta_{-} \\
  \delta_{+} & {\rm if} \: \mathcal{F}_{k,e} > \delta_{+} \\
  \mathcal{F}_{k,e} & {\rm otherwise}
 \end{cases}
\end{equation}
where $\delta_{-} = \max(0, x_{k, e} - \mu)$, $\delta_{+} = \min(1, x_{k, e} + \mu)$, for the fixed move limit $\mu\in(0,1)$ and
\begin{equation}
 \label{eq:OCupdate2}
 \mathcal{F}_{k,e} = x_{k, e}
 \left( - \frac{\partial_{e} c_{k}}
 {\tilde{\lambda}_{k}\partial_{e} V_{k}}\right)^{1/2}
\end{equation}
depends on the element sensitivities.

In \eqref{eq:OCupdate2} $\tilde{\lambda}_{k}$ is the approximation to the current Lagrange multiplier $\lambda^{\ast}_{k}$ associated with the volume constraint. This is obtained by imposing $V(\hat{\mathbf{x}}_{k+1}(\tilde{\lambda})) - f|\Omega_{h}| \approx 0$, e.g. by bisection on an interval $\Lambda^{(0)}_{k} \supset \lambda^{\ast}_{k}$.

\section{Matlab implementation and speedups}
 \label{Sec:codeStructure}

The Matlab routine for 2D problems (see \autoref{Sec:2DcodeComplianceMinimization}) is called with the following arguments
\begin{lstlisting}[basicstyle=\scriptsize\ttfamily,breaklines=true,numbers=none,frame=single]
top99neo(nelx,nely,volfrac,penal,rmin,ft,ftBC,eta,beta,move,maxit);
\end{lstlisting}
where \texttt{nelx}, \texttt{nely} define the physical dimensions and the mesh resolution, \texttt{volfrac} is the allowed volume fraction on the overall domain (i.e. $\mathcal{A}\cup \mathcal{P}$), \texttt{penal} the penalization used in \eqref{eq:simpPenalization} and \texttt{rmin} the filter radius for \eqref{eq:filter}. The parameter \texttt{ft} is used to select the filtering scheme: density filtering alone if \texttt{ft=1}, whereas \texttt{ft=2} or \texttt{ft=3} also allows the projection \eqref{eq:projection}, with \texttt{eta} and \texttt{beta} as parameters. \texttt{ftBC} specifies the filter boun dary conditions (\texttt{'N'} for zero-Neumann or \texttt{'D'} for zero-Dirichlect), \texttt{move} is the move limit used in the OC update and \texttt{maxit} sets the maximum number of redesign steps

The routine is organized in a set of operations which are performed only once and the loop for the TO iterative re--design. The initializing operations are grouped as follows
\begin{scriptsize}
 \[
  \begin{aligned}
   &\texttt{PRE.1) MATERIAL AND CONTINUATION PARAMETERS} \\
   &\texttt{PRE.2) DISCRETIZATION FEATURES}\\
   &\texttt{PRE.3) LOADS, SUPPORTS AND PASSIVE DOMAINS} \\
   &\texttt{PRE.4) DEFINE IMPLICIT FUNCTIONS} \\
   &\texttt{PRE.5) PREPARE FILTER} \\
   &\texttt{PRE.6) ALLOCATE AND INITIALIZE OTHER PARAMETERS}\\
  \end{aligned}
 \]
\end{scriptsize}
and below we give details only about parameters and instructions not found in the \texttt{top88} code.

To apply continuation on the generic parameter ``$\texttt{par}$'', a data structure is defined
\begin{small}
\[
 \texttt{parCont = \{istart, maxPar, isteps, deltaPar\};}
\]
\end{small}
such that the continuation starts when \texttt{loop=istart} and the parameter is increased by \texttt{deltaPar} each \texttt{isteps}, up to the value \texttt{maxPar}. This is implemented in Line 6 and 7 for the penalization parameter $p$ and the projection factor $\beta$, respectively. The update is then performed, by the instruction (see Line 92)
\begin{lstlisting}[basicstyle=\scriptsize\ttfamily,breaklines=true,numbers=none,frame=single]
   par=par+(loop>=parCnt{1}).*(par<parCnt{2}).*mod(loop,parCnt{3})==0).*parCnt{4}
\end{lstlisting}
making use of compact logical operations. Continuation can be switched off e.g. by setting $\texttt{maxPar<=par}$, or $\texttt{istart>=maxit}$.

The blocks defining the discretization (\texttt{PRE.2)}) contain some changes compared to \texttt{top88}. The number of elements (\texttt{nEl}), DOFs (\texttt{nDof}) and the set of node numbers (\texttt{nodeNrs}) are defined explicitly, to ease and shorten some following instructions. The setup of indices \texttt{iK}, \texttt{jK}, used for the sparse assembly, is performed in Lines 15-21 and follows the concept detailed in \autoref{sSec:speedupAssembly}. The coefficients of the lower diagonal part of the elemental stiffness matrix are defined in vectorized form, such that $\texttt{Ke} = \mathcal{V}(K^{(s)}_{e})$ (see Lines 22--26). \texttt{Ke} is used for the assembly strategy described in \autoref{sSec:speedupAssembly}. However, in Lines 27--29 we also recover the complete elemental matrix (\texttt{Ke0}), used to perform the double product $\mathbf{u}^{T}_{e}K_{e}\mathbf{u}_{e}$ when computing the compliance sensitivity \eqref{eq:sensitivityComplianceVolume}. Altough this could also be written in terms of the matrix $K^{(s)}_{e}$ only, this option would increase the number of matrix/vector multiplications.

In \texttt{PRE.3)} the user can specify the set of restrained (\texttt{fixed}) and loaded (\texttt{lcDof}) DOFs and passive regions ($\mathcal{P}_{1} \leftrightarrow \texttt{pasS}$ and $\mathcal{P}_{0} \leftrightarrow \texttt{pasV}$) for the given configuration. Supports and loads are defined as in the \texttt{top88} code, whereas passive domains may be specified targeting a set of column and rows from the array \texttt{elNrs}. Independently of the particular example, Lines 34--36 define the vector of applied loads, the set of free DOFs, and the sets of active $\mathcal{A} \leftrightarrow \texttt{act}$ design variables.

In order to make the code more compact and readable, operations which are repeatedly performed within the TO optimization loop are defined through inline functions in \texttt{PRE.4)} (Lines 38--43). The filter operator is built in \texttt{PRE.5)} making use of the built--in Matlab function \texttt{imfilter}, which represents a much more efficient alternative to the explicit construction of the neighboring array. A similar approach was already outlined by \cite{andreassen-etal_11a}, pointing to the Matlab function \texttt{conv2}, which is however not completely equivalent to the original operator, as it only allows zero--Dirichlect boundary conditions for the convolution operator. Here we choose \texttt{imfilter}, which is essentially as efficient as \texttt{conv2}, but gives the flexibility to specify zero--Dirichlect (default option), or zero--Neumann boundary conditions.

Some final initializations and allocations are performed in \texttt{PRE.6)}. The design variables are initialized with the modified volume fraction, accounting for the passive domains (Line 52--53) and the constant volume sensitivity \eqref{eq:sensitivityComplianceVolume} is computed in Line 51.

Within the redesign loop, the following five blocks of operations are repeatedly performed
\begin{scriptsize}
 \[
  \begin{aligned}
   &\texttt{RL.1) COMPUTE PHYSICAL DENSITY FIELD} \\
   &\texttt{RL.2) SETUP AND SOLVE EQUILIBRIUM EQUATIONS} \\
   &\texttt{RL.3) COMPUTE SENSITIVITIES} \\
   &\texttt{RL.4) UPDATE DESIGN VARIABLES AND APPLY CONTINUATION} \\
   &\texttt{RL.5) PRINT CURRENT RESULTS AND PLOT DESIGN} \\
  \end{aligned}
 \]
\end{scriptsize}

In block \texttt{RL.1)}, the physical field is obtained, applying the density filter and, if selected, also the projection. If \texttt{ft=3}, the special value of the threshhold $\texttt{eta}$ giving a volume--preserving projection is computed, as discussed in \autoref{sSec:filtersVP}.

\begin{figure}
 \centering
  \fbox{
   \includegraphics[scale = 0.25, keepaspectratio]
   {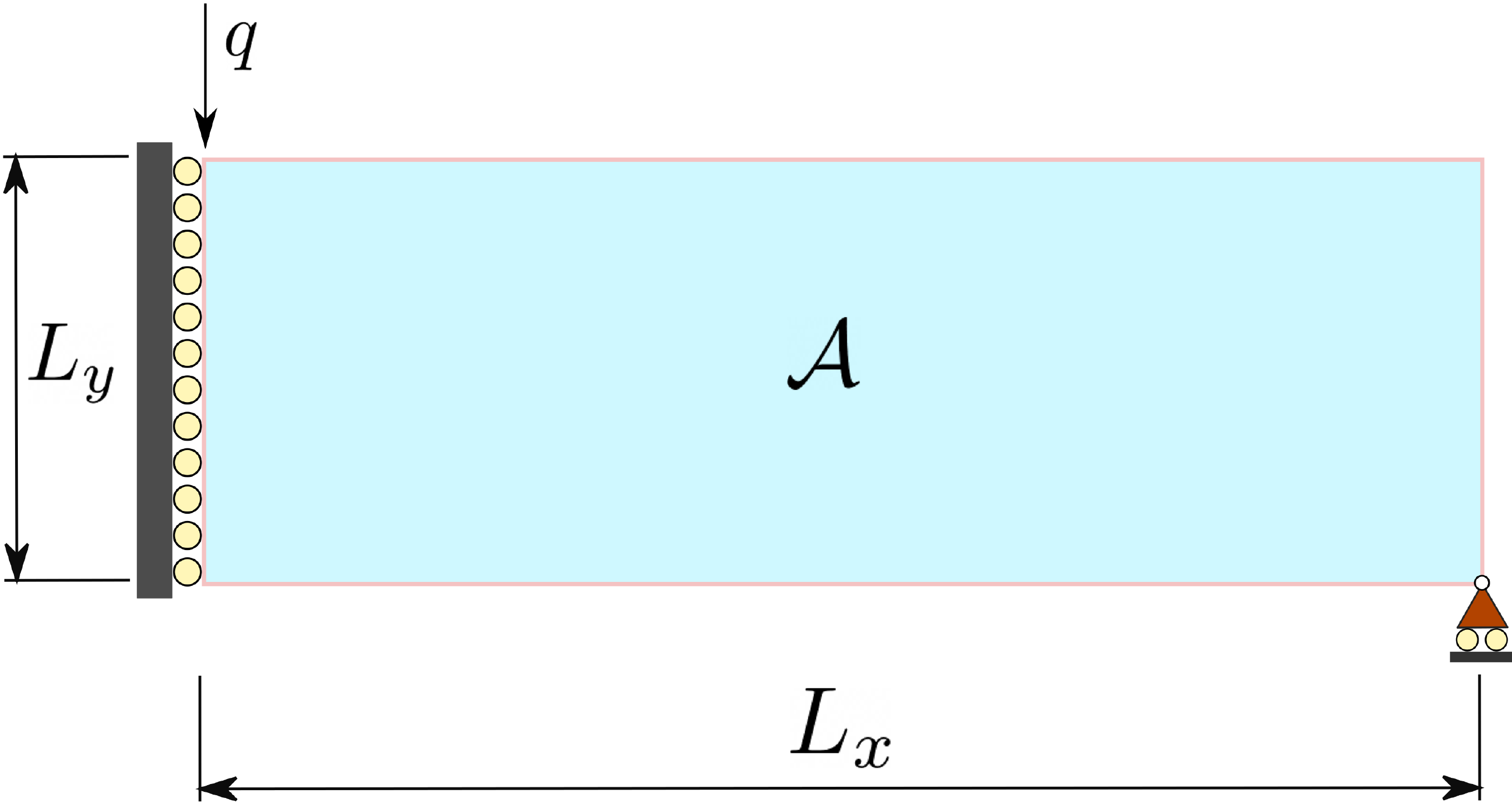}}
 \caption{\small{Geometrical setting for the MBB example}}
 \label{fig:sketchExamples}
\end{figure}

The stiffness interpolation and its derivative (\texttt{sK}, \texttt{dsK}) are defined and the stiffness matrix is assembled (see Lines 73--76). Ideally, one could also get rid of Lines 73-74 and directly define \texttt{sK} in Line 75 and \texttt{dsK} within Line 79. However, we decide to keep these operations apart, enhanching the readability of the code and to ease the specification of different interpolation schemes. \autoref{eq:stateProblem} is solved on Line 77 by using the Matlab function \texttt{decomposition}, which can work with only half of the stiffness matrix (see \autoref{sSec:speedupAssembly}). The sensitivity of compliance is computed and the backfiltering operations \eqref{eq:sensitivityComplianceVolumeDV} are performed in \texttt{RL.3)}.

The update \eqref{eq:OCupdate}, with the nested application of the bisection process for finding $\tilde{\lambda}_{k}$, is implemented in \texttt{RL.4)} (Lines 86--91) and we remark that \texttt{lm} represents $\sqrt{\lambda}$.

Some information about the process are printed and the current design is plotted in \texttt{RL.5)} (Lines 94--97). On small discretizations repeated plotting operations absorb a significant fraction of the CPU time (e.g. $15\%$ for $m = 4800$). Therefore, one might just plot the final design, moving Lines 96--97 outside the redesign loop.

Tests in the following have been run on a laptop equipped with an Intel(R) Core(TM) i7-5500U@2.40GHz CPU, 15GB of RAM and Matlab 2018b running in serial mode under Ubuntu 18.04 (but a similar performance is expected in Windows setups). We will often refer to the half MBB beam example (see \autoref{fig:sketchExamples}) for numerical testing. Unless stated otherwise we choose $\Omega_{h} = 300\times 100$, $f = 0.5$ and $r_{\rm min} = 8.75$ \citep{sigmund_07a}. The load, having total magnitude $|q| = 1$ is applied to the first node. No passive domains are introduced for this example, therefore \texttt{pasS=[];}, \texttt{pasV=[];} and we set $E_{1} = 1$, $E_{0} = 10^{-9}$ and $\nu = 0.3$ in all the tests.

\begin{table*}[tb]
 \caption{\small{Number of entries in the array $\mathcal{I}$ and corresponding memory requirement for the 2D and 3D test discretizations. White background refers to the {\rm F} strategy with coefficients specified as \texttt{double}, cyan background to the {\rm H} strategy and light green to the {\rm H} strategy and element specified as \texttt{int32}. The {\rm H} strategy cuts $|\mathcal{I}|$ and memory of $\approx 44\%$ in 2D and $\approx 48\%$ in 3D. Then, specifying the indexes as \texttt{int32} further cuts memory of another $50\%$}}
 \label{tab:indexingforAssemblage}
 \centering
  \begin{tabular}{c|c|ccccc}
   \hline\noalign{\smallskip}
   \multirow{6}{*}{2D} &                              $m$ & $120^{2}$ & $240^{2}$ & $480^{2}$ & $960^{2}$ & $1920^{2}$ \\
    \noalign{\smallskip}\hline
                       & \multirow{2}{*}{$|\mathcal{I}|$} & $1,843,200$ & $7,372,800$ & $29,491,200$ & $117,964,800$ & $-$ \\
                       &                                  & \cellcolor{LightCyan}{$1,036,800$} & \cellcolor{LightCyan}{$4,147,200$} & \cellcolor{LightCyan}{$16,588,800$} &  \cellcolor{LightCyan}{$66,355,200$} & \cellcolor{LightCyan}{$265,420,800$} \\
                       &                                  &      &    &     &         \\
                       &     \multirow{3}{*}{memory (MB)} & 14.7 & 59 & 235 & 943 & - \\
                       &                                  &  \cellcolor{LightCyan}{8.3} & \cellcolor{LightCyan}{33.2} & \cellcolor{LightCyan}{132.7} & \cellcolor{LightCyan}{530.8} & \cellcolor{LightCyan}{2123} \\
                       &                                  &  \cellcolor{lime}{4.1} & \cellcolor{lime}{16.6} & \cellcolor{lime}{66.3} & \cellcolor{lime}{265.4} & \cellcolor{lime}{1061} \\
    \noalign{\smallskip}\hline
                       &                                  &      &    &     &         \\
   \multirow{6}{*}{3D} &                              $m$ &   $8^{3}$ &  $16^{3}$ & $32^{3}$ & $64^{3}$ & $128^{3}$ \\
    \noalign{\smallskip}\hline
                       & \multirow{2}{*}{$|\mathcal{I}|$} & $589,824$ & $4,718,592$ & $37,748,736$ &  $301,898,888$ & $2,415,919,104$ \\
                       &                                  & \cellcolor{LightCyan}{$307,200$} & \cellcolor{LightCyan}{$2,457,200$} & \cellcolor{LightCyan}{$19,660,800$} & \cellcolor{LightCyan}{$157,286,400$} & \cellcolor{LightCyan}{$1,258,291,200$} \\
                       &                                  &      &    &     &         \\
                       &     \multirow{3}{*}{memory (MB)} & 4.7 & 37.7 & 302 & 2416 & 9664 \\
                       &                                  & \cellcolor{LightCyan}{2.5} & \cellcolor{LightCyan}{19.7} & \cellcolor{LightCyan}{157.3} & \cellcolor{LightCyan}{1258} & \cellcolor{LightCyan}{5033} \\
                       &                                  & \cellcolor{lime}{1.2} & \cellcolor{lime}{9.8} & \cellcolor{lime}{78.6} & \cellcolor{lime}{629.1} & \cellcolor{lime}{2516} \\
   \noalign{\smallskip}\hline
  \end{tabular}
\end{table*}

\subsection{Speedup of the assembly operation}
 \label{sSec:speedupAssembly}
 
In $\texttt{top88}$ the assembly of the global stiffness matrix is performed by the built--in Matlab function $\texttt{sparse}$
\begin{lstlisting}[basicstyle=\scriptsize\ttfamily,breaklines=true,numbers=none,frame=single]
 K = sparse( iK, jK, sK );
 K = ( K +K' ) / 2;
\end{lstlisting}
where $\texttt{sK}\in\mathbb{R}^{m*d^{2}\times 1}$ collects the coefficients of all the elemental matrices in a column--wise vectorized form (i.e. $\mathcal{V}(K_{e})$) and $\texttt{iK}$, $\texttt{jK}$ are the sets of indices mapping each $\texttt{sK(i)}$ to the global location $\texttt{K(iK(i),jK(i))}$.

These two sets are set up through the operations
\begin{equation}
 \label{eq:formalBuildIKJK}
  \texttt{iK} = \mathcal{V}\left[(C \otimes \mathbf{1}_{d})^{T}\right] \ , \qquad
  \texttt{jK} = \mathcal{V}\left[(C \otimes \mathbf{1}^{T}_{d})^{T}  \right]
\end{equation}
where $C_{[m \times d]}$ is the connectivity matrix and ``$\otimes$'' is the Kronecker product \citep{book:horn-johnson}. The size of the array $\mathcal{I} = [ \texttt{iK}, \texttt{jK} ] \in\mathbb{N}^{m*d^{2}\times 2}$ grows very quickly with the number of elements $m$, especially for 3D discretizations (see \autoref{tab:indexingforAssemblage}), and even though its elements are integers, the \texttt{sparse} function requires them to be specified as \texttt{double} precision numbers. The corresponding memory burden slows down the assembly process and restricts the size of problems workable on a laptop.

The efficiency of the assembly can be substantially improved by
\begin{enumerate}
 \item Acknowledging the symmetry of both $K_{e}$ and $K$;
 \item Using an assembly routine working with \texttt{iK} and \texttt{jK} specified as integers;
\end{enumerate}

To understand how to take advantage of the symmetry of matrices, we refer to \autoref{fig:settingFigures}(b) and to the connectivity matrix $C$. Each coefficient $C_{ej}\in\mathbb{N}$ addresses the global DOF targeted by the $j$--th local DOF of element $e$. Therefore \eqref{eq:formalBuildIKJK} explicitly reads
\begin{equation}
 \label{eq:indexingExplicit}
 \begin{aligned}
  \texttt{iK}^{e} & =
  \{\underbrace{\mathbf{c}_{e}, \mathbf{c}_{e}, \ldots, \mathbf{c}_{e}}_{d \ {\rm times}} \} \\
  \texttt{jK}^{e} & = \{
  \underbrace{c_{e1}, \ldots, c_{e1}}_{d \ {\rm times}}, 
  \underbrace{c_{e2}, \ldots, c_{e2}}_{d \ {\rm times}},
  \ldots,
  \underbrace{c_{ed}, \ldots, c_{ed}}_{d \ {\rm times}}\}
 \end{aligned}
\end{equation}
where $\mathbf{c}_{e} = \left\{ c_{e1}, c_{e2}, \ldots, c_{ed} \right\}$ is the row corresponding to element $e$.

If we only consider the coeffcients of the (lower) symmetric part of the elemental matrix $K^{(s)}_{e}$ and their locations into the global one $K^{(s)}$, the set of indices can be reduced to
\begin{small}
\begin{equation}
 \label{eq:reducedIndices}
  \begin{aligned}
   \texttt{iK}^{e} & = \{
   c_{e1}, \ldots, c_{ed}, c_{e2}, \ldots, c_{ed}, \ldots,
   c_{e3}, \ldots, c_{ed}, \ldots, c_{ed} \} \\
   \texttt{jK}^{e} & = \{
   \underbrace{c_{e1}, \ldots, c_{e1}}_{d \ {\rm times}}, 
   \underbrace{c_{e2}, \ldots, c_{e2}}_{(d-1) \ {\rm times}}, 
   \underbrace{c_{e3}, \ldots, c_{e3}}_{(d-2) \ {\rm times}}, 
   \ldots, c_{ed} \}
  \end{aligned}
\end{equation}
\end{small}
and the overall indexing array becomes $\mathcal{I}_{r} = [\texttt{iK},\texttt{jK}] \in \mathbb{N}^{\tilde{d}*m\times 2}$ where $\tilde{d} = \sum^{d}_{j=1} \sum_{i\leq j} i$. The entries of the indexing array and the memory usage are reduced by approx. $45\%$ (see \autoref{tab:indexingforAssemblage}).

The set of indices \eqref{eq:reducedIndices} can be constructed by the following instructions (see Lines 15--21)
\begin{lstlisting}[basicstyle=\scriptsize\ttfamily,breaklines=true,numbers=none,frame=single]
 [ setI, setII ] = deal( [ ] );
 for j=1:8
    setI=cat(2,setI,j:8);
    setII = cat(2,setII,repmat(j,1,8-j+1));
 end
 [iK , jK] = deal(cMat(:, sI)', cMat(:, sII)');
 Iar = sort([iK(:),jK(:)], 2,'descend'); clear iK jK
\end{lstlisting}
which can be adapted to any isoparametric 2D/3D element just by changing accordingly the number \texttt{d} of elemental DOFs. In the attached scripts, based on 4--noded bilinear $Q4$ and 8--noded trilinear $H8$ elements, we set \texttt{d=8} and \texttt{d=24}, respectively. The last instruction sorts the indices as $\texttt{iK(i)} > \texttt{jK(i)}$, such that $K^{(s)}$ contains only sub--diagonal terms.

The syntax \texttt{K=sparse(iK,jK,sK)} now returns the lower triangular matrix $K^{(s)}$ and we remark that the full operator can be recovered by
\begin{equation}
 \label{eq:recoverFullMatrix}
  K = K^{(s)} + (K^{(s)})^{T} - {\rm diag}[K^{(s)}]
\end{equation}
which costs as much as the averaging operation $\frac{1}{2}(K+K^{T})$, performed in \texttt{top88} to get rid of roundoff errors. However, the Matlab built--in Cholesky solver and the corresponding \texttt{decomposition} routine can use just $K^{(s)}$, if called with the option \texttt{'lower'}.

\begin{figure}
 \centering
  \subfloat[2D discretization]{
   \includegraphics[scale = 0.45, keepaspectratio]
    {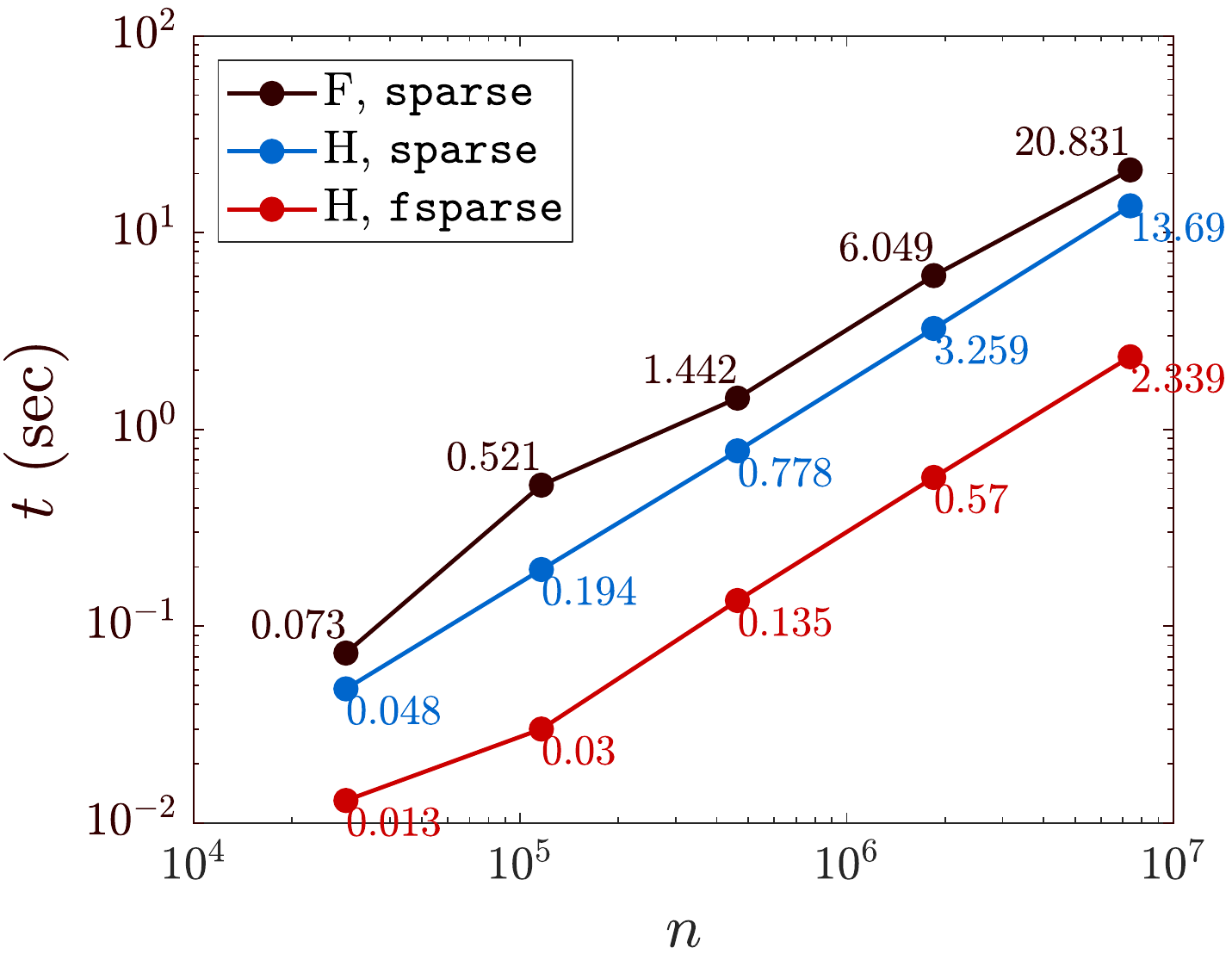}} \\
  \subfloat[3D discretization]{
   \includegraphics[scale = 0.45, keepaspectratio]
    {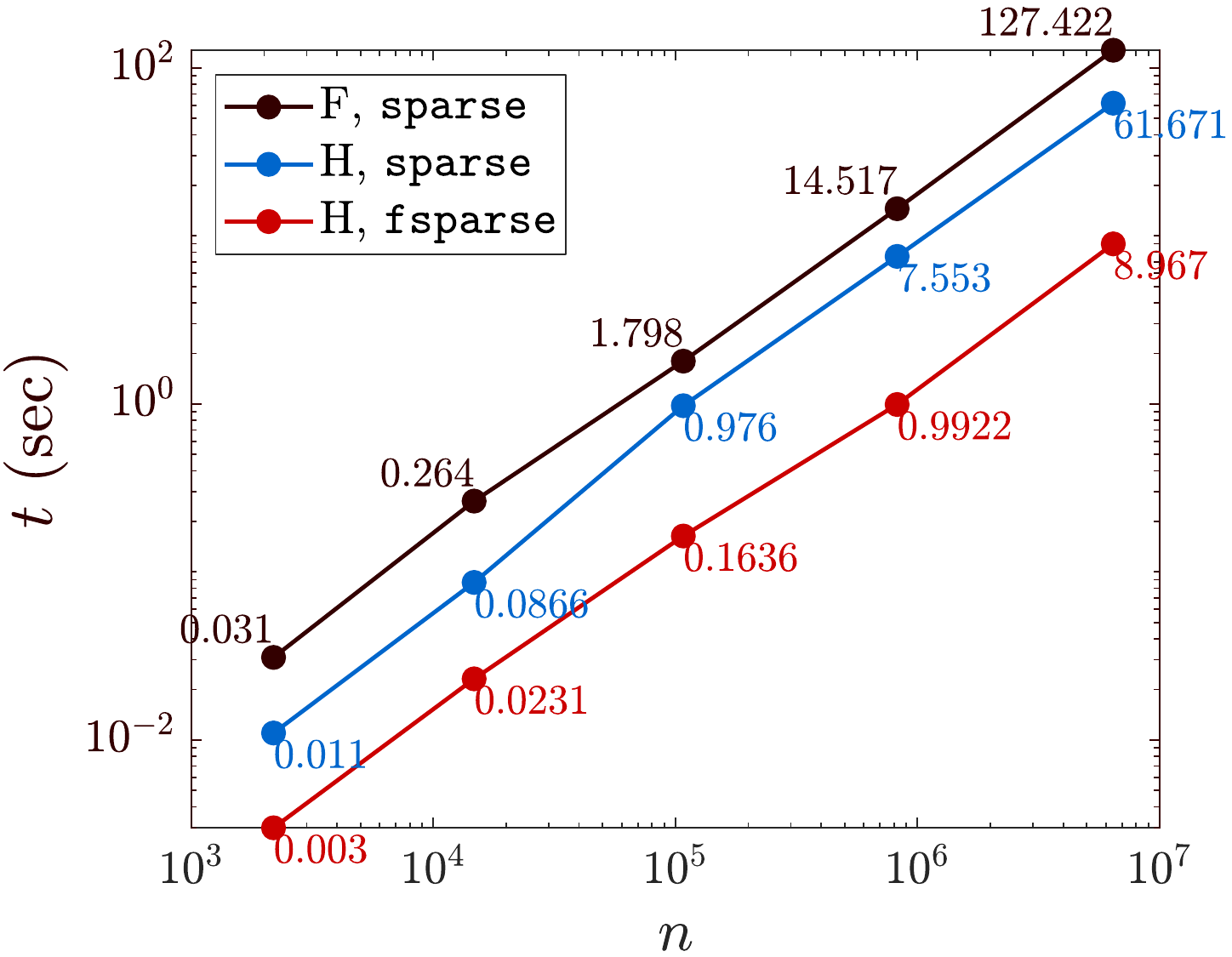}}
 \caption{\small{Scaling of assembly time performed with the 3 strategies discussed in \autoref{sSec:speedupAssembly}. Compared to the standard ({\rm F}) assembly, the {\rm H} strategy alone cuts near $50\%$ of time and memory, and with the use of \texttt{fsparse} gives an overall efficiency improvement of 10--15 times}}
 \label{fig:scalingAssemblage}
\end{figure}

Point 2 gives the most dramatic improvement, and can be accomplished by using routines developed by independent researchers. The \texttt{sparse2} function, from Suite Sparse \citep{suitesparse}, was already pointed out by \citep{andreassen-etal_11a} as a better alternative to the built--in Matlab \texttt{sparse}; however, no quantitative comparisons were performed. According to the CHOLMOD reference manual \citep{davis_09a}, \texttt{sparse2} works exactly as \texttt{sparse}, but allowing the indices \texttt{iK} and \texttt{jK} to be specified as integers (accomplished by defining this type for the connectivity matrix, see Lines 11 and 13).

Here we suggest the ``\texttt{fsparse}'' routine, developed by \cite{engblom-lukarski_16a}. Besides working with integers \texttt{iK} and \texttt{jK}, the function enhanches the efficiency of the sparse assembly by a better sorting of the operations. From our experience on a single core process, \texttt{fsparse} gives a speedup of 170--250$\%$ compared to \texttt{sparse2}, and is also highly parallelizable \citep{engblom-lukarski_16a}. Defining the sets \texttt{iK} and \texttt{jK} as \texttt{int32} type, we can drastically cut the memory requirements, still representing $n \approx 2.1\cdot 10^{9}$ numbers, far beyond the size of problems one can tackle in Matlab.

In order to use \texttt{fsparse}, one needs to download the ``\texttt{stenglib}'' library\footnote{https://github.com/stefanengblom/stenglib} and follow the installation instructions in the \texttt{README.md} file. The packages of the library can be installed by running the ``\texttt{makeall.m}'' file. As \texttt{fsparse} is contained within the folder ``Fast'', one may only select this folder when running \texttt{makeall.m}.

We test the efficiency of the assembly approaches on 2D and 3D uniform discretizations with $m^{2}$ and $m^{3}$ elements, respectively. \autoref{fig:scalingAssemblage} shows time scalings for the different strategies: {\rm ``F''} corresponds to the assembly in \texttt{top88}, {\rm ``H''} takes advantage of the matrix symmetry only and {\rm ``H,\texttt{fsparse}''} correponds to the use of the \texttt{fsparse} routine \citep{engblom-lukarski_16a} also. All the approaches exibith a linear scaling of CPU time w.r.t the DOFs number. However, half the CPU time can be cut just by assembling $K^{(s)}$ (strategy {\rm H,\texttt{sparse}}). Therefore, we definitely recommend this to users who aim to solve medium--size ($10^{5}$ to $10^{6}$ DOFs) structural TO problems on a laptop. However, the most substantial savings follow from using \texttt{fsparse} \citep{engblom-lukarski_16a} and by coupling these two strategies ({\rm H,\texttt{fsparse}}) speedups of 10 for the 2D and 15 for 3D setting can be achieved. It is worth to highlight that a 3D stiffness matrix of the size of $\approx 9\cdot 10^{5}$ can be assembled in less than a second and even one of size $6.2\cdot 10^{6}$ can be assembled on a laptop in less than 10s. For this last case the sole storage of the arrays \texttt{iK}, \texttt{jK} and \texttt{sK} would cause a memory overflow, ruling out the {\rm ``F''} approach.

\subsection{Speedup of the OC update}
 \label{sSec:filtersVP}
 
The cost of the re--design step $\mathbf{x}_{k+1} = \mathcal{U}(\mathbf{x}_{k})$ is proportional to the number of bisections  ($n_{\rm bs}$) required for computing the approximation $\tilde{\lambda}_{k} \approx \lambda^{\ast}_{k}$. The following estimate \citep{book:quarteroni-etal_00}
\begin{equation}
 \label{eq:bisectEstimate}
 n_{\rm bs} \geq \frac{\log(|\Lambda^{(0)}|) - \log(\tau)}{\log(2)} - 1
\end{equation}
is a lower bound to this number for a given accuracy $\tau > |\lambda^{\ast}_{k} - \tilde{\lambda}_{k}|$ and it is clear that $n_{\rm bs}$ would decrease if $\Lambda^{(0)}$, the initial guess for the interval bracketing $\lambda^{\ast}_{k}$, could be shrunk. Moreover, the volume constraint should be imposed on the physical field ($\tilde{\mathbf{x}}$ or $\hat{\mathbf{x}}$) and, in the original \texttt{top88} implementation, this requires a filter application at each bisection step, which may become expensive. 

The efficiency of the re--design step can be improved by a two step strategy
\begin{enumerate}
 \item Using volume--preserving filtering schemes;
 \item Estimating the interval $\Lambda^{(0)}_{k}$ bracketing the current Lagrange multiplier $\lambda^{\ast}_{k}$
\end{enumerate}

Concerning point 1, the density filter is naturally volume--preserving (i.e. $V(\mathbf{x}_{k}) = V(\tilde{\mathbf{x}}_{k})$) \citep{bourdin_01a,bruns-tortorelli_01a}. Therefore, the volume constraint can be enforced on $V(\mathbf{x}_{k})$ as long as the density filter alone is considered (\texttt{ft=1}). The relaxed Heaviside projection \eqref{eq:projection}, on the other hand, is not volume--preserving for any $\eta$; thus it would require one filter--and--projection application at each bisection step. However, \eqref{eq:projection} can also be made volume--preserving by computing, for each $\tilde{\mathbf{x}}_{k}$, the threshhold $\eta^{\ast}_{k}$ such that \citep{xu-etal_10a,li-khandelwal_15b}
\begin{equation}
 \label{eq:optEtaProblem}
 \eta^{\ast}_{k} \longrightarrow \min_{\eta \in [0, 1]}
 |V(\hat{\mathbf{x}}_{k}(\eta))-V(\tilde{\mathbf{x}}_{k})|
\end{equation}

This can be done, e.g. by the Newton method, starting from the last computed $\eta^{\ast}_{k-1}$ and provided the derivative of \eqref{eq:projection} with respect to $\eta$
\begin{equation}
 \label{eq:derivativeHeavisideEta}
  \frac{\partial V(\tilde{\mathbf{x}}(\eta))}{\partial \eta} =
 - 2 \beta \sum_{i\in\mathcal{A}}
 \frac{(e^{\beta(1-\tilde{x}_{i})}-e^{\beta(\tilde{x}_{i}-1)})(e^{\beta x}-e^{\beta \tilde{x}_{i}})}
 {(e^{\beta}-e^{-\beta})[e^{\beta(\tilde{x}_{i}-\eta)}+e^{\beta(\eta-\tilde{x}_{i})}]^{2}}
\end{equation}

Existence of $\eta^{\ast} \in [0, 1]$ for all $\tilde{\mathbf{x}} \in [0, 1]^{m}$ follows from the fact that $g( \eta ) = V(\hat{\mathbf{x}}(\eta))-V(\tilde{\mathbf{x}})$ is continuous on $[0,1]$ and $g(0)g(1) \leq 0$; uniqueness follows from the fact that $\frac{\partial g}{\partial \eta} < 0$ for all $\eta \in (0,1)$.

\begin{figure}
 \centering
  \subfloat[Dashed lines show the cumulative number of Newton iterations]{
   \includegraphics[scale = 0.45, keepaspectratio]
    {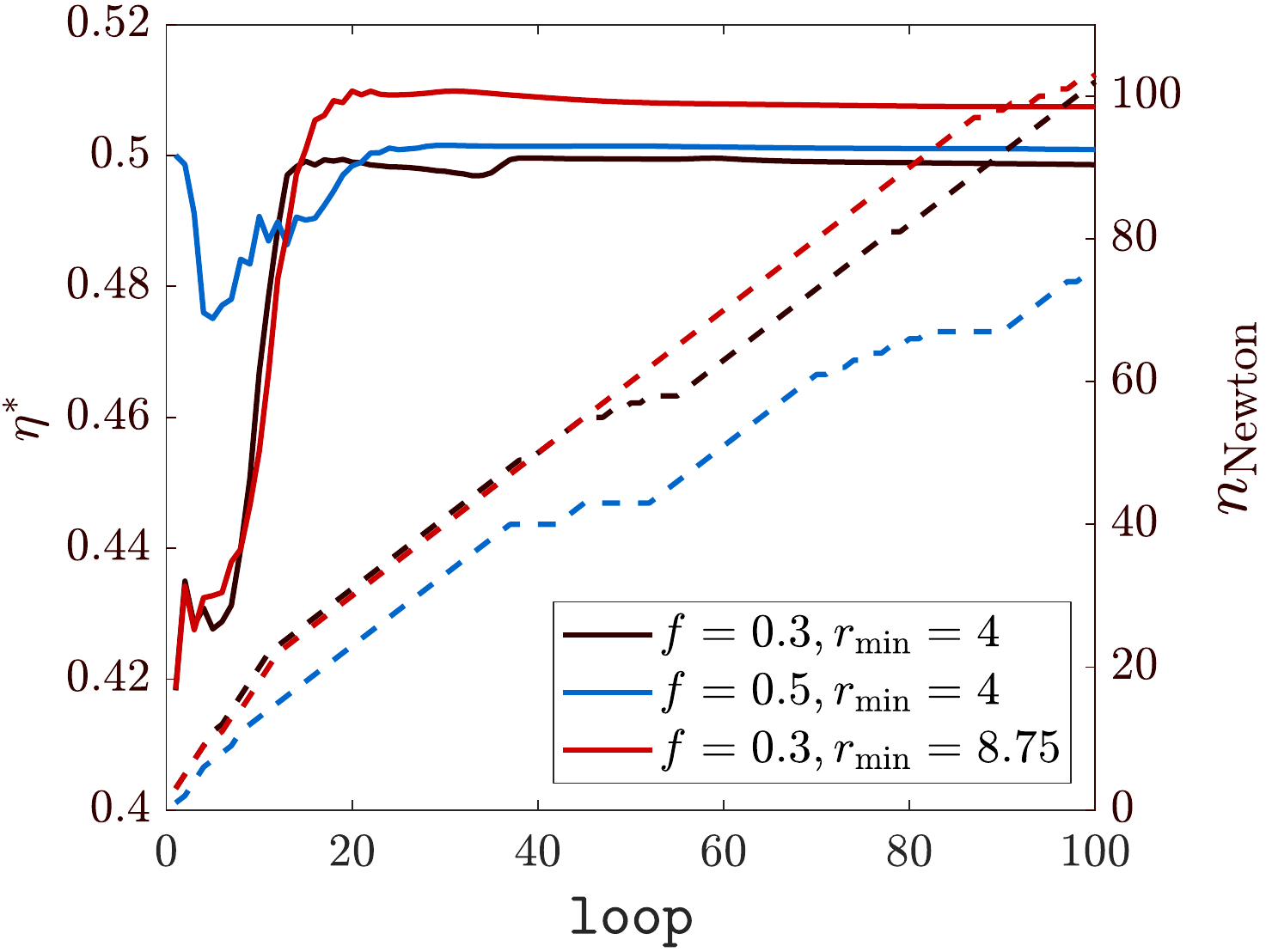}} \\
  \subfloat[Dashed lines show the cumulative number of bisection steps. Green lines refer to the use of the exiplicit primal-dual iteration discussed in \autoref{Sec:lmEstimate} for computing $\lambda^{\ast}$]{
   \includegraphics[scale = 0.45, keepaspectratio]
    {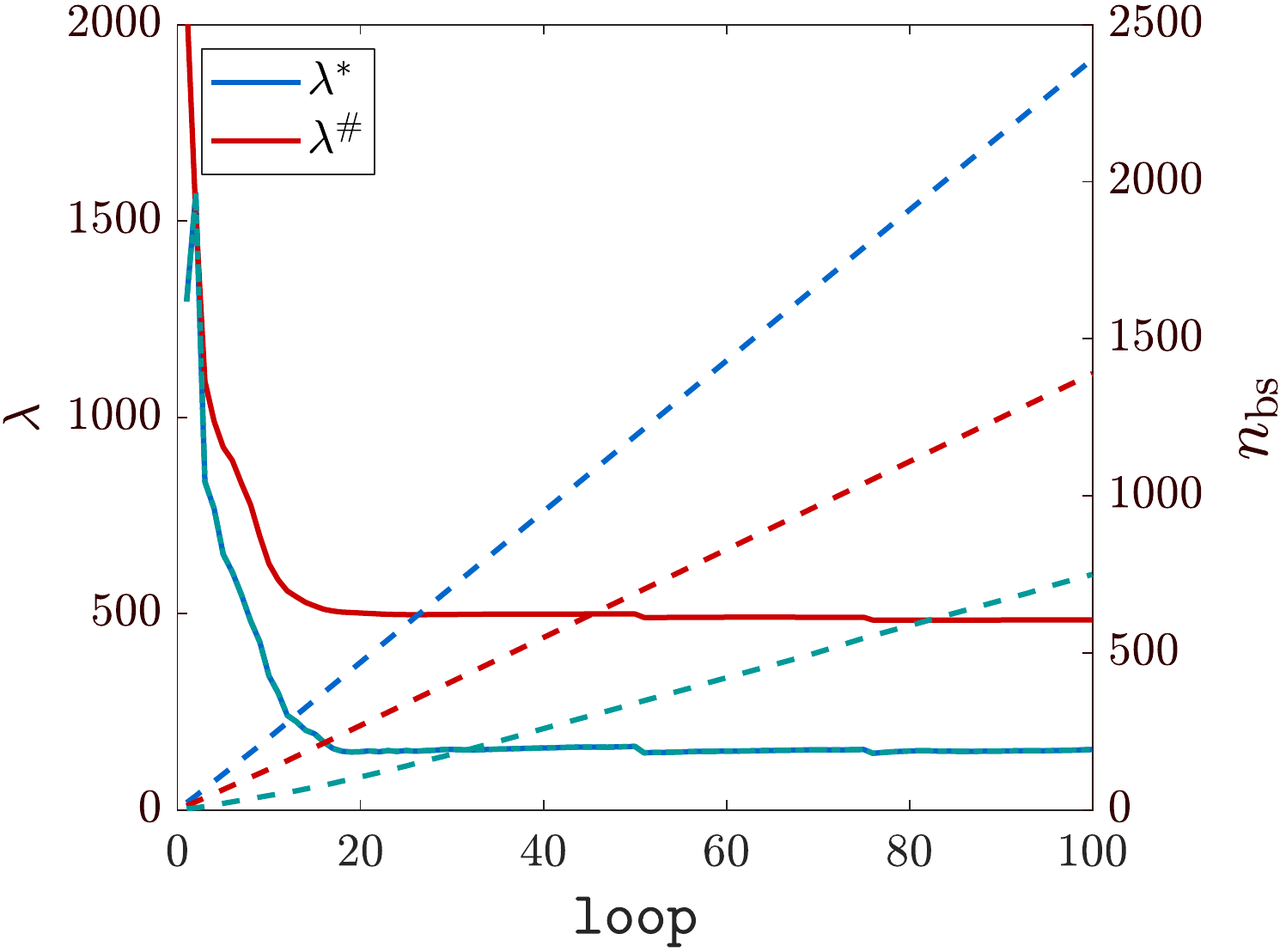}}
 \caption{\small{Evolution of the parameter $\eta^{\ast}$ realizing the equivalence $V(\tilde{\mathbf{x}}) = V(\hat{\mathbf{x}})$, for different volume fractions $f$ and filter radii $r_{\rm min}$ (a) and evolution of the Lagrange multiplier estimate $\lambda^{\#}$ given by \eqref{eq:estimateLM} compared to $\lambda^{\ast}$ (b). For both plots, the cumulative number of Newton iterations $n_{\rm Newton}$ (viz. number of bisection steps $n_{\rm bs}$) is shown against the right axis}}
 \label{fig:optEtaandlambda}
\end{figure}

Numerical tests on the MBB beam show that generally $\eta^{\ast}_{k}\in \left[ 0.4, 0.52\right]$, the larger variability occurring for low volume fractions (see \autoref{fig:optEtaandlambda} (a)). We also observe that $\eta^{\ast}_{k}$ takes values slightly above 0.5 when $r_{\rm min}$ is increased or $\beta$ is raised. Convergence to $\eta^{\ast}_{k}$ is generally attained in 1--2 Newton iterations (see \autoref{fig:optEtaandlambda} (a)).

The procedure for computing $\eta^{\ast}_{k}$ from \eqref{eq:optEtaProblem}, with tolerance $\epsilon = 10^{-6}$ and initial guess $\eta_{0} = \texttt{eta}$, provided by the user, is implemented in Lines 63--67, that are executed if the routine $\texttt{top99neo}$ is called with the parameter \texttt{ft=3}. Otherwise, if \texttt{ft=2}, the input threshhold \texttt{eta} is kept fixed. In case of the latter, the volume constraint should be consistently applied on $V(\hat{\mathbf{x}})$, otherwise some violation or over--shooting of the constraint will happen. In particular, if the volume constraint is imposed on $\mathbf{x}$ and $\eta$ is kept fixed, one has $V(\hat{\mathbf{x}}) > f |\Omega_{h}|$, if $\eta < 0.5$, and $V(\hat{\mathbf{x}}) < f |\Omega_{h}|$, if $\eta > 0.5$.

Even tough we usually oberved small differences, these may result in local optima or bad designs, especially for low volume fractions or high $\beta$ values. Therefore, accounting for this more general situation Lines 87--91 should be replaced by the following
\begin{lstlisting}[basicstyle=\scriptsize\ttfamily,breaklines=true,numbers=none,frame=single]
while (l(2)-l(1))/(l(2)+l(1))>1e-4
  lmid=0.5*(l(1)+l(2));
  x=max(max(min(min(ocP/lmid,xU),1),xL),0);
  if ft > 0
    xf=imfilter(reshape(x,nely,nelx)./Hs,h);
    [xf(pasS),xf(pasV)]=deal(1,0);
    if ft>1, xf=prj(xf(:),eta,beta); end
  end
  if mean(xf(:))>volfrac,l(1)=lmid;else,l(2)=lmid;end
end 
\end{lstlisting}

However, there could be other situations when one cannot rely on volume preserving filters (e.g. when imposing length scale through robust design). Therefore, a more general strategy to reduce the cost of the OC update is to cut the number of bisection steps.

To this end, the selection of the initial bracketing interval $\Lambda^{(0)}_{k}$ may build upon the upper bound estimate for $\lambda^{\ast}_{k}$ (\cite{hestenes_69a,arora-etal_91a})
\begin{equation}
 \label{eq:estimateLM}
  \lambda^{\#}_{k} = \left[ \frac{1}{mf}
  \sum^{m}_{e=1} x_{k,e}
  \left(
  -\frac{\partial_{e}c_{k}}
  {\partial_{e}V_{k}}
  \right)^{1/2}
  \right]^{2}
\end{equation}

More details on the derivation of \eqref{eq:estimateLM} are given in \autoref{Sec:lmEstimate}. The behavior of the estimate \eqref{eq:estimateLM} is shown in \autoref{fig:optEtaandlambda}(b) for the MBB example. The overall number of bisections ($n_{bs}$) in order to compute $\lambda^{\ast}_{k}$ meeting the tolerance $\tau = 10^{-8}$ when considering $\Lambda^{(0)}_{k} = [0, \lambda^{\ast}_{k}]$ is cut by about $50\%$, compared with the one required by starting from $\Lambda^{(0)}=[0,10^{9}]$) as in \texttt{top88}. Moreover, if no projection is applied, \eqref{eq:estimateLM} could be used together with \eqref{eq:OCupdate} to perform an explicit Primal-Dual iteration to compute $(\mathbf{x}_{k+1}, \lambda^{\ast}_{k})$ and this would reduce the number of steps even more (see green curve in \autoref{fig:optEtaandlambda}(b)).

However, in the basic versions of the codes, given in \autoref{Sec:2DcodeComplianceMinimization} and \autoref{Sec:3DcodeComplianceMinimization}, we consider the bisection process and \eqref{eq:estimateLM} is used to bracket the search interval, as this procedure is more general.

\subsection{Acceleration of the OC iteration}
 \label{sSec:OCacceleration}

The update rule \eqref{eq:OCupdate} resembles a Fixed--Point (FP) iteration $\mathbf{x}_{k+1} = \mathcal{U}(\mathbf{x}_{k})$, generating a sequence $\{\mathbf{x}_{k}\}$ converging to a point such that $\mathbf{r} = \mathcal{U}(\mathbf{x}^{\ast})-\mathbf{x}^{\ast} = \mathbf{0}$.

Several methods are available to speedup the convergence of such a sequence \citep{brezinski-chehab_98a,ramiere-helfer_15a}, somehow belonging to the family of quasi--Newton methods \citep{eyert_96a}. The acceleration proposed by \cite{anderson_65a}, for instance, is nowadays experiencing a renewed interest \citep{fang-saad_09a,pratapa-etal_16a,peng-etal_18a}, and has recently been applied to TO by \cite{thesis:li-paulino_18}.

Anderson acceleration takes into account the residuals $\mathbf{r}_{i}$, their differences $\Delta\mathbf{r}_{i} = \mathbf{r}_{i+1}-\mathbf{r}_{i}$ and the differences of the updates $\Delta\mathbf{x}_{i} = \mathbf{x}_{i+1}-\mathbf{x}_{i}$ for the last $m_{r}$ iterations (i.e. $i=k-m_{r}, \ldots, k-1$), and obtains the new element of the vector sequence as
\begin{equation}
 \label{eq:andersonUpdateStep}
  \mathbf{x}_{k+1} =
  \mathbf{x}^{\#}_{k} + \zeta\mathbf{r}^{\#}_{k}
\end{equation}
where $\zeta\in [0, 1]$ is a damping coefficient and
\begin{equation}
 \begin{aligned}
  \mathbf{x}^{\#}_{k} & =
  \mathbf{x}_{k} - \sum^{k-1}_{i=k-m_{r}}\gamma^{(k)}_{i}
  \Delta \mathbf{x}_{i} =
  \mathbf{x}_{k} - X_{k} \boldsymbol{\gamma}_{k} \\
  \mathbf{r}^{\#}_{k} & =
  \mathbf{r}_{k} - \sum^{k-1}_{i=k-m_{r}}\gamma^{(k)}_{i}
  \Delta \mathbf{r}_{i} =
  \mathbf{r}_{k} - R_{k} \boldsymbol{\gamma}_{k} \\
 \end{aligned}
\end{equation}

The coefficients $\gamma^{(k)}_{i}$ minimize the following
\begin{equation}
 \label{eq:leastSquaresProblem}
  \{\gamma^{(k)}_{i}\}^{m_{r}}_{i=1} \rightarrow
  \min_{\boldsymbol{\gamma}}
  \|\mathbf{r}^{\#}_{k}(\boldsymbol{\gamma})\|^{2}_{2}
\end{equation}

The rationale behind the method is to compute a rank--$m_{r}$ update of the inverse Jacobian matrix $J^{-1}_{k}$ of the nonlinear system $\mathbf{r}_{k} = \mathbf{0}$. This has been shown to be equivalent to a multi--secant Broyden method \citep{eyert_96a,fang-saad_09a} starting from $J^{-1}_{0} = -\zeta I$.

The update rule \eqref{eq:andersonUpdateStep} is usually applied only once each $q$ steps. Thus we can write more generally $\mathbf{x}_{k+1} = \mathbf{x}_{k} + \mathbf{z}_{k}$, where \citep{pratapa-etal_16a}
\begin{equation}
 \label{eq:paeUpdating}
 \mathbf{z}_{k} = 
 \begin{cases}
  \alpha \mathbf{r}_{k}  & {\rm if} \: \frac{k+1}{q} \notin \mathbb{N} \\
  \zeta I - ( X_{k} + \zeta F_{k} ) \boldsymbol{\gamma}_{k} &
  {\rm if } \: \frac{k+1}{q} \in \mathbb{N} \\
 \end{cases}
\end{equation}
($\alpha\in (0,1)$) obtaining the so--called Periodic Anderson Extrapolation (PAE) \citep{pratapa-etal_16a,thesis:li-paulino_18}.

The implementation can be obtained, e.g. by adding the following few lines after the OC step (Line 91)
\begin{lstlisting}[basicstyle=\scriptsize\ttfamily,breaklines=true,numbers=none,frame=single]
fres = x( act ) - xT( act );
 if loop >= q0
    sel = mod(loop - q0, q)==0;
    mix = sel*xi + alpha*(1-sel);
    x(act) = pae(xT(act),fres,mr,loop-q0,mix,q);
    x(x > 1) = 1; x(x < 0) = 0;
 end
% --------------------------
function [ xnew ] = pae( x, r, m, it, mix, q )
persistent X R Xold Rold;  dp = 0;
if it > 1
   k = mod( it - 1, m ) + 1;
   [X(:, k), R(:, k)] = deal(x - Xold, r - Rold);
   if rem( it-1, q ) == 0
       dp = (X+mix*R)* ((R'*R)\(R'*r));
   end
end
xnew = x + mix * r - dp; Xold = x; Rold = r;
end
\end{lstlisting}
where the part solving \eqref{eq:leastSquaresProblem} and the update has been put in a separate routine for better efficiency.

In the above we use the ``$\backslash$'' for solving the least squares problem \eqref{eq:leastSquaresProblem}; however, strategies based on a QR (or SVD) decomposition may be preferred in terms of numerical stability. We refer to \cite{fang-saad_09a} for a deeper discussion on this point.

In order to assess the effect of different filtering schemes and the introduction of parameter continuation, Anderson acceleration is tested on the MBB example considering the following options
\begin{itemize}
 \item[T1] Density filter alone, $p = 3$;
 \item[T2] Density--and--projection filter, with $\eta^{\ast}$ computed from \eqref{eq:optEtaProblem} and $\beta=2$;
 \item[T3] As T2, but with continuation on both $\beta$ and $p$, defined by the parameters \texttt{betaCnt=\{250,16,25,2\}} and \texttt{penalCnt=\{50,3,25,0.25\}};
 \item[T4] As T2, but for the discretization $\Omega_{h} = 600\times 200$
\end{itemize}

For all the cases, the TO loop stops when $\|\mathbf{r}_{k}\|_{2}/\sqrt{m} < 10^{-6}$, where the residual is defined with respect to the physical variables (i.e. $\mathbf{r}_{k} = \tilde{\mathbf{x}}_{k} - \tilde{\mathbf{x}}_{k-1}$ for T1 and $\mathbf{r}_{k} = \hat{\mathbf{x}}_{k} - \hat{\mathbf{x}}_{k-1}$ for T2--T4). The acceleration is applied each $q = 4$ steps, considering the last $m_{r} = 4$ residuals, starting from iteration $q_{0} = 20$ for T1--T2 and from $q_{0} = 500$ for T3--T4, when both continuations have finished. We set $\alpha = 0.9$ for the non--accelerated steps. The choice $m_{r} = 4$ is based on the observation that convergence improvements increase very slowly for $m_{r}>3$ (\cite{anderson_65a,eyert_96a}). However, a deeper discussion about the influence of all parameters on the convergence is outside the scope of the present work and we refer to \cite{thesis:li-paulino_18} or, in a more general context, to \cite{walker-ni_11a} for this.

\begin{small}
\begin{table}
 \caption{\small{Comparison of convergence--related parameters for the standard (T) and accelerated (T-${\rm PAE}$) TO tests, for the MBB example}}
 \label{tab:comparisonOCPAE}
 \centering
  \begin{tabular}{l|cccccc}
   \hline\noalign{\smallskip}
      & it. & $c$ & $\Delta c$ & $\|\mathbf{r}\|_{2}/\sqrt{m}$ & $m_{ND}$ \\
   \noalign{\smallskip}\hline
   T1 & $2500$ & $252.7$ & $4.2 \cdot 10^{ -8}$ & $1.03 \cdot 10^{-5}$ & $0.025$ \\
   T1--PAE &     828 & $258.9$ & $4.2 \cdot 10^{-10}$ & $9.95 \cdot 10^{-7}$ & $0.021$ \\
   \noalign{\smallskip}\hline
   T2 & $2500$ & $246.1$ & $5.1 \cdot 10^{ -8}$ & $3.21 \cdot 10^{-5}$ & $0.023$ \\
   T2--PAE &     352 & $253.9$ & $6.2 \cdot 10^{ -9}$ & $9.97 \cdot 10^{-7}$ & $0.014$ \\
   \noalign{\smallskip}\hline
   T3 & $2500$ & $199.6$ & $1.1 \cdot 10^{ -4}$ & $1.91 \cdot 10^{-3}$ & $0.014$ \\
   T3--PAE &     752 & $197.5$ & $3.7 \cdot 10^{ -8}$ & $8.72 \cdot 10^{-7}$ & $0.007$ \\
   \noalign{\smallskip}\hline
   T4 & $2500$ & $191.8$ & $2.0 \cdot 10^{ -7}$ & $3.21 \cdot 10^{-5}$ & $0.006$ \\
   T4--PAE &     818 & $192.1$ & $2.5 \cdot 10^{ -7}$ & $9.97 \cdot 10^{-7}$ & $0.001$ \\
   \noalign{\smallskip}\hline
  \end{tabular}
\end{table}
\end{small}

\begin{figure}
 \centering
  \subfloat[T1]{
   \includegraphics[scale = 0.3, keepaspectratio]
    {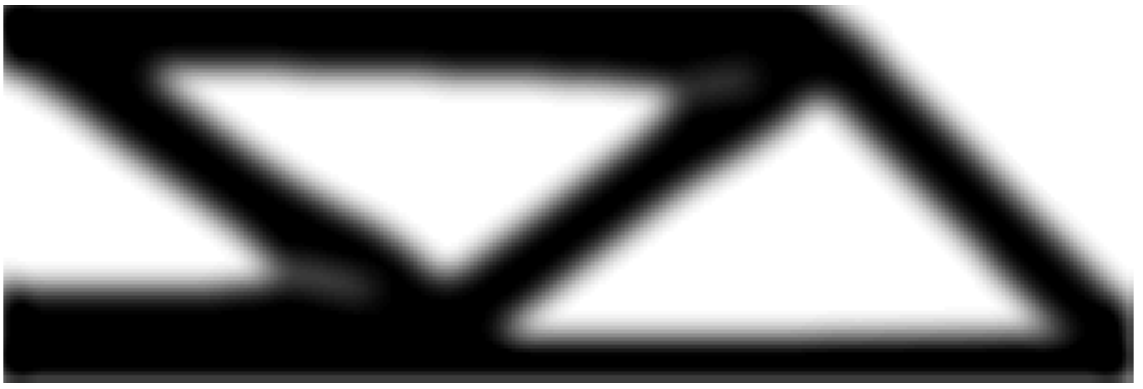}} \qquad
  \subfloat[T1-PAE]{
   \includegraphics[scale = 0.3, keepaspectratio]
    {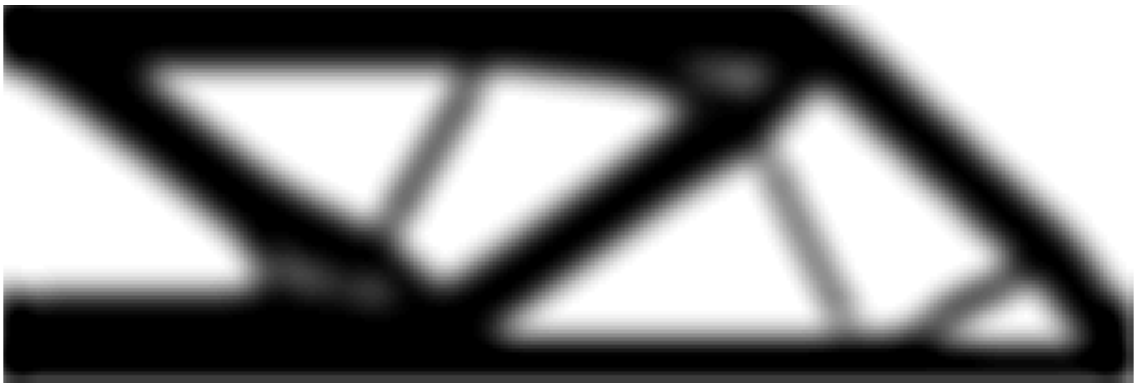}} \\
  \subfloat[T2]{
   \includegraphics[scale = 0.3, keepaspectratio]
    {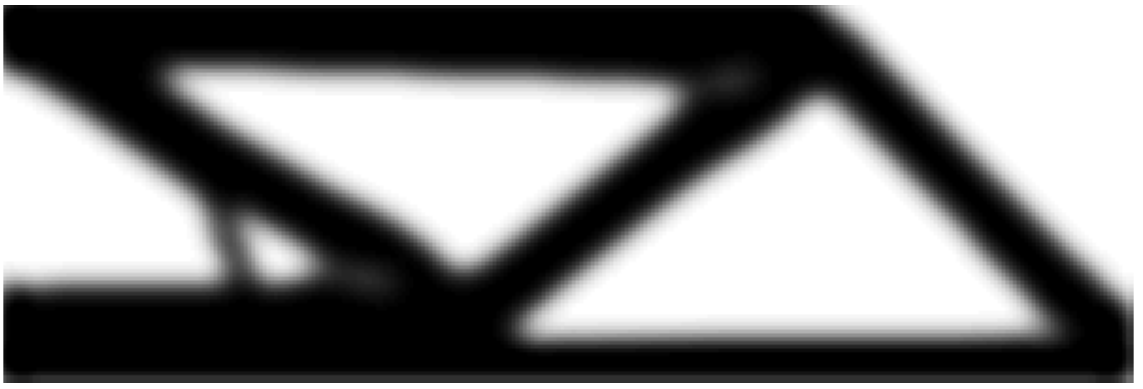}} \qquad
  \subfloat[T2-PAE]{
   \includegraphics[scale = 0.3, keepaspectratio]
    {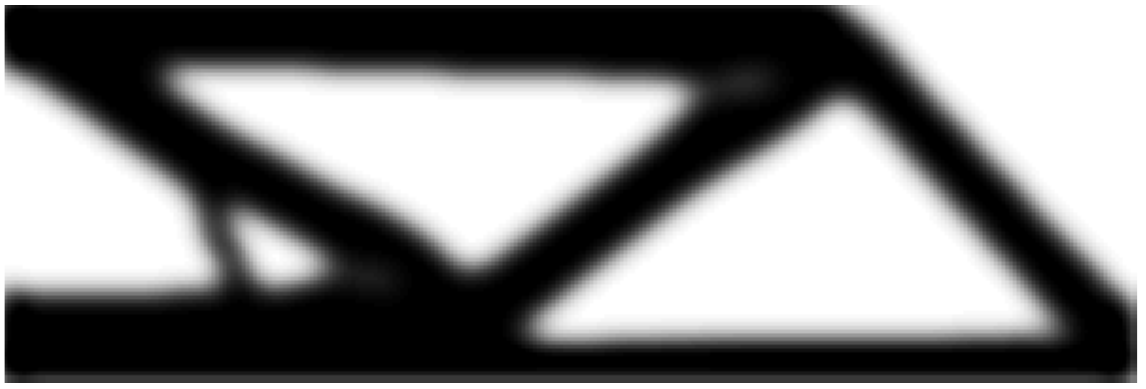}} \\
  \subfloat[T3]{
   \includegraphics[scale = 0.3, keepaspectratio]
    {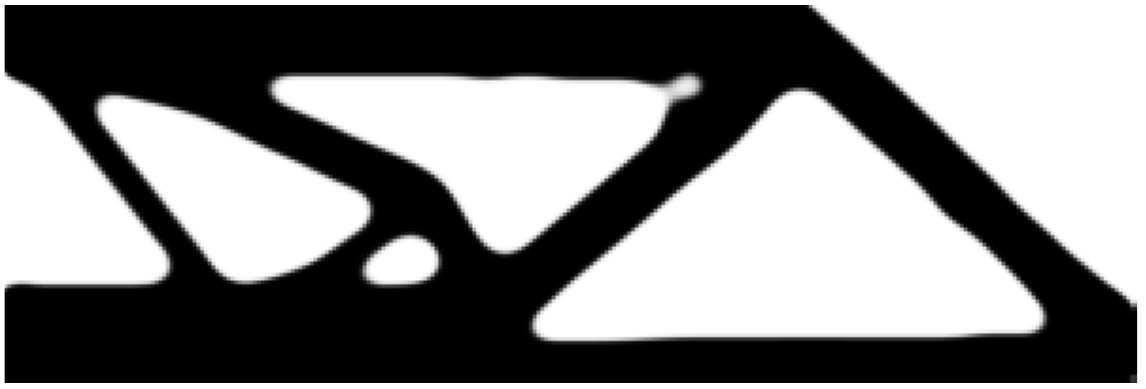}} \qquad
  \subfloat[T3-PAE]{
   \includegraphics[scale = 0.3, keepaspectratio]
    {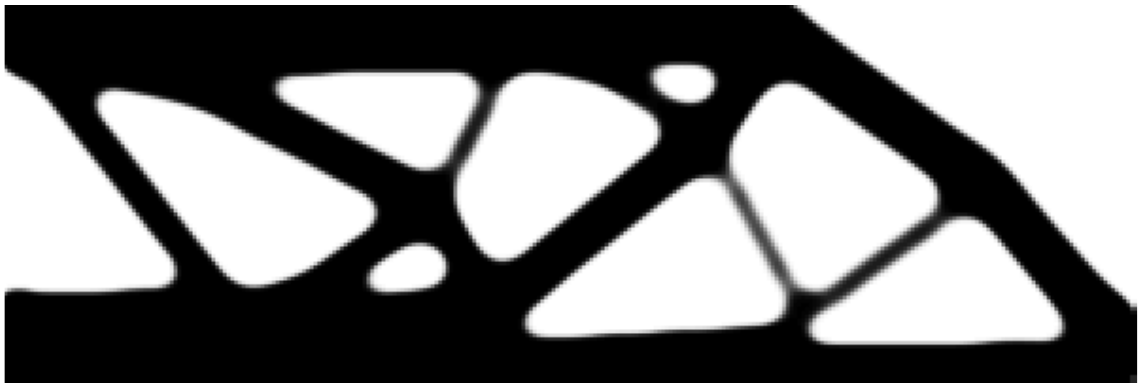}} \\
  \subfloat[T4]{
   \includegraphics[scale = 0.3, keepaspectratio]
    {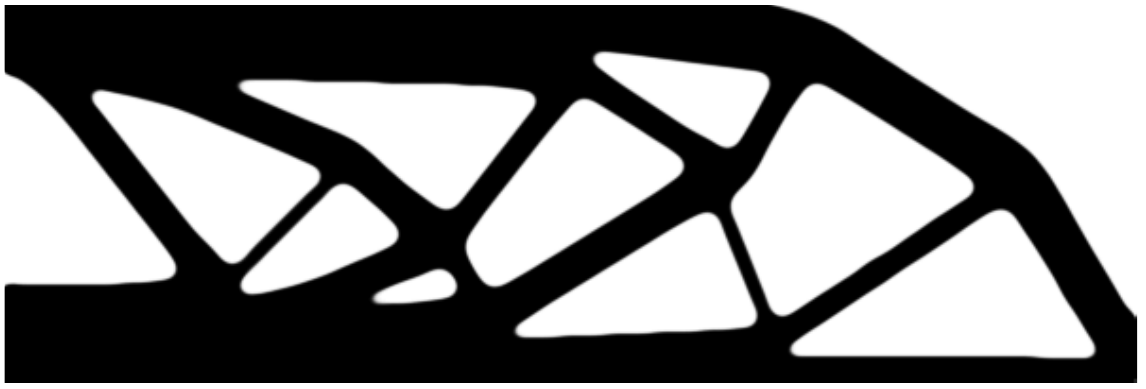}} \qquad
  \subfloat[T4-PAE]{
   \includegraphics[scale = 0.3, keepaspectratio]
    {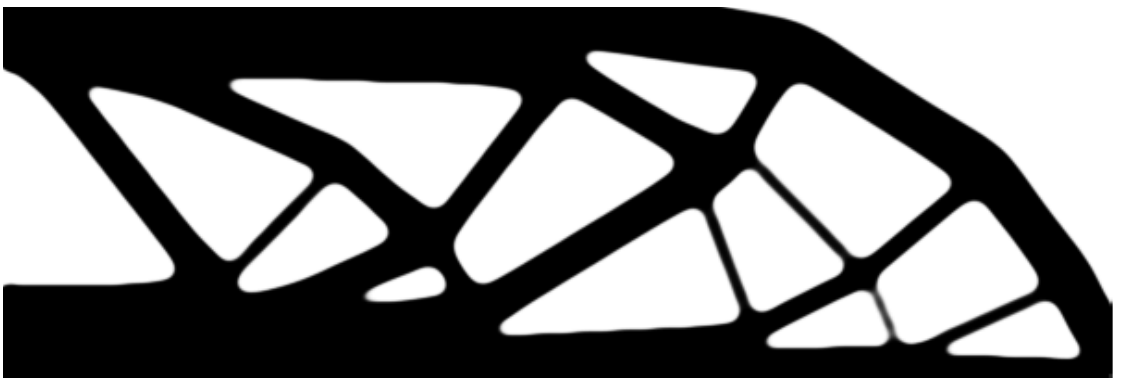}}
 \caption{\small{Optimized designs obtained without (left column) and with Anderson acceleration (right column) of the TO loop}}
 \label{fig:testPAEtop}
\end{figure}

\begin{figure*}
 \centering
 \subfloat[T1]{
   \includegraphics[scale = 0.25, keepaspectratio]
    {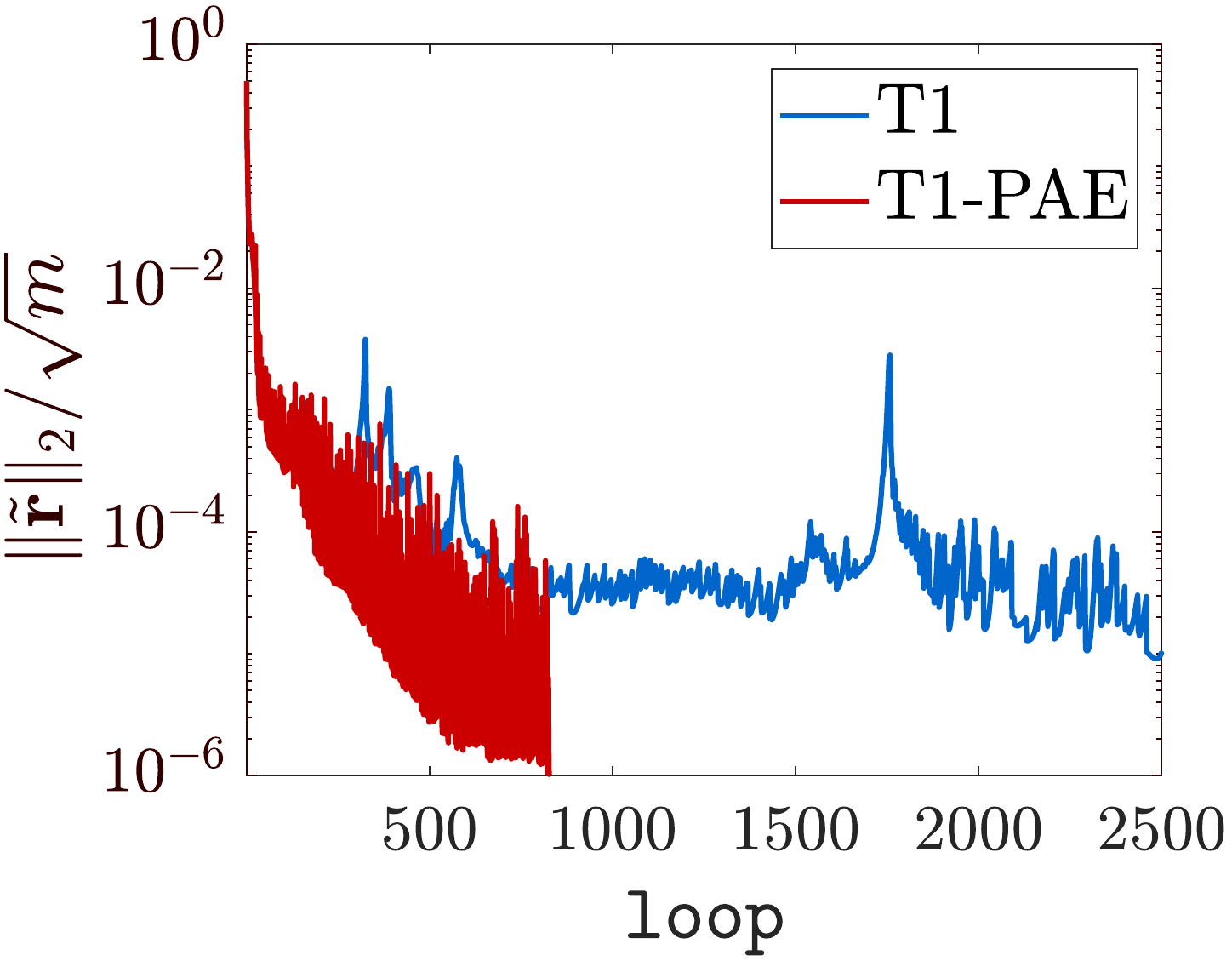}} \quad
  \subfloat[T2]{
   \includegraphics[scale = 0.25, keepaspectratio]
    {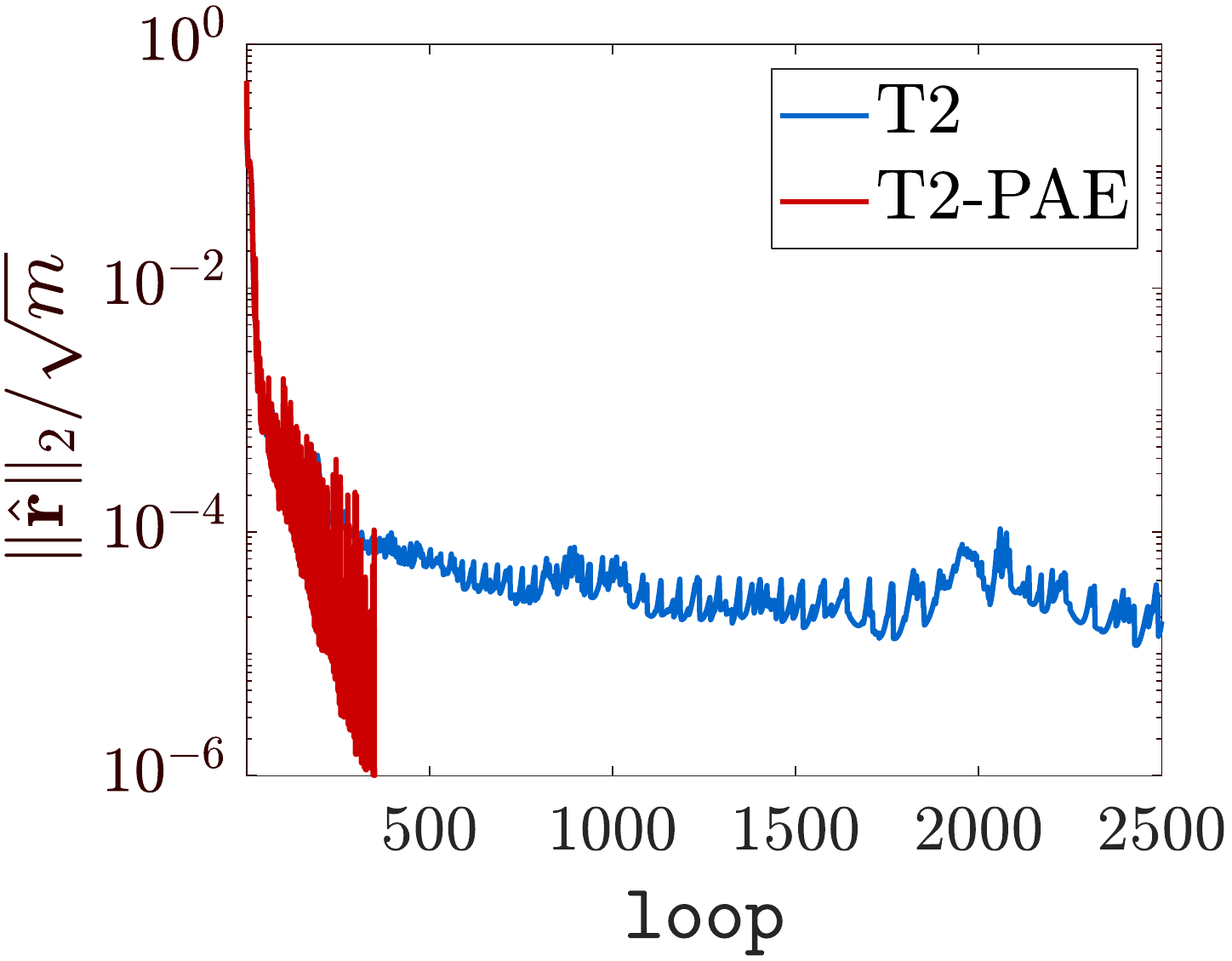}} \quad
  \subfloat[T3]{
   \includegraphics[scale = 0.25, keepaspectratio]
    {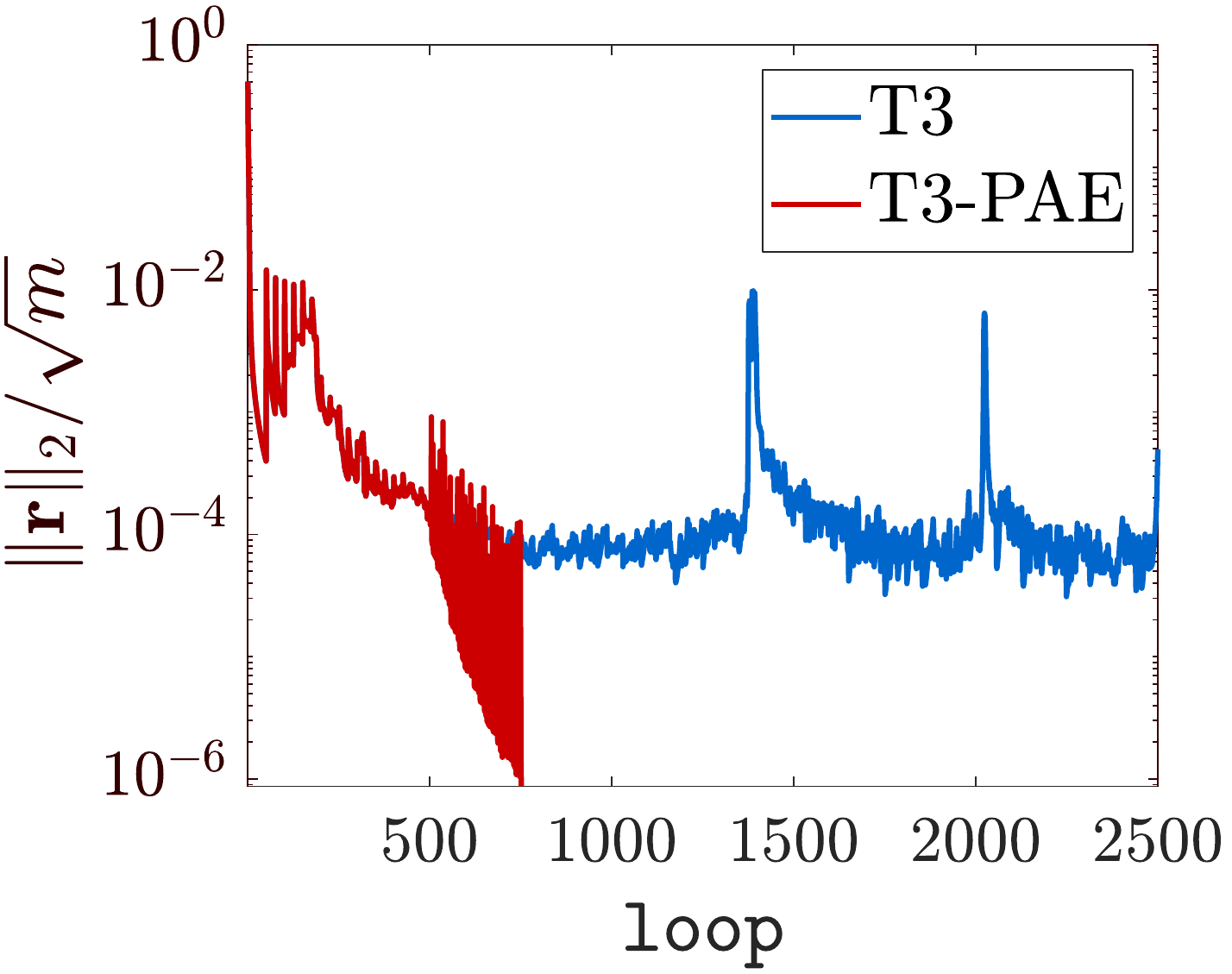}} \quad
  \subfloat[T4]{
   \includegraphics[scale = 0.25, keepaspectratio]
    {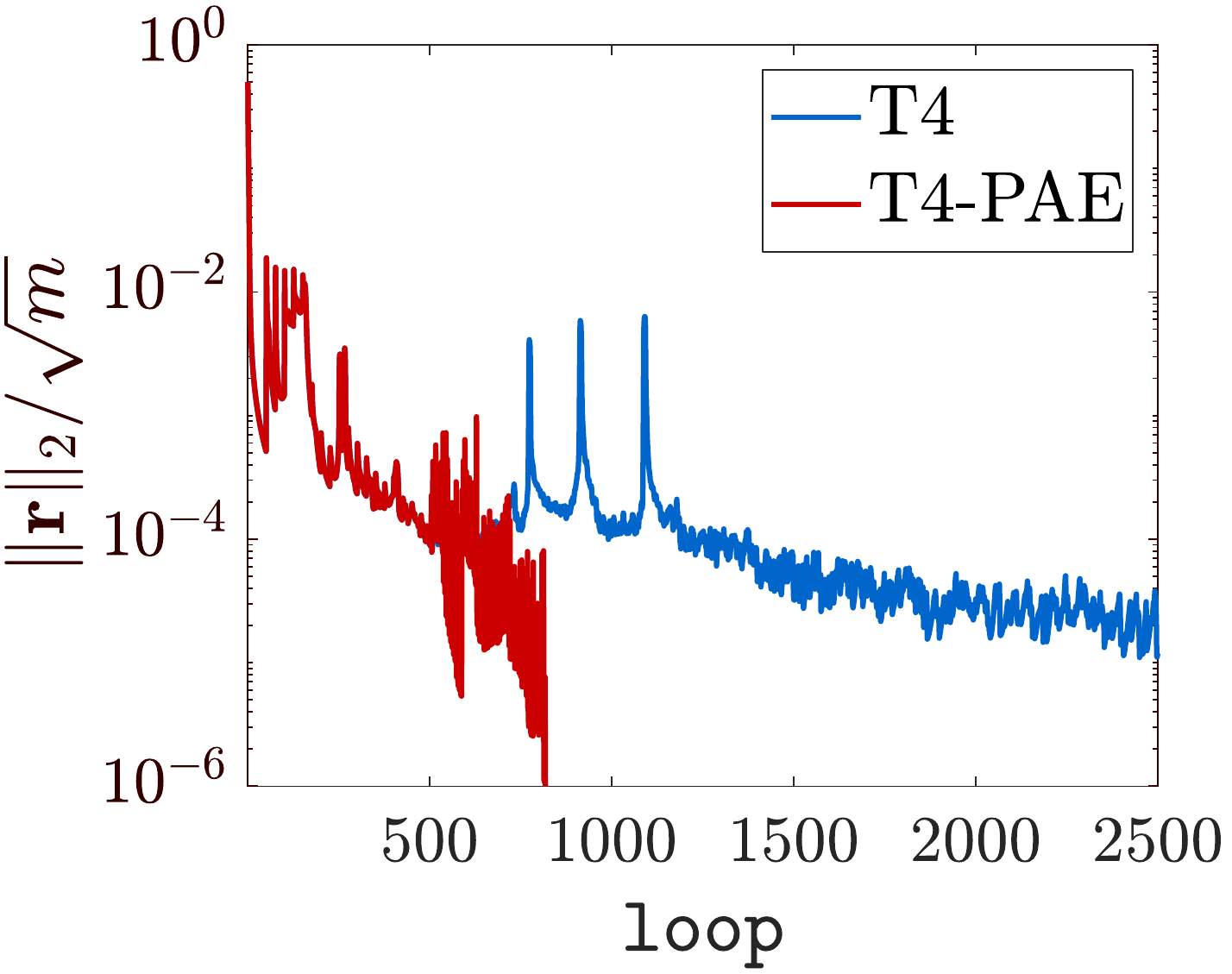}}
  \\
  \subfloat[T1]{
   \includegraphics[scale = 0.25, keepaspectratio]
    {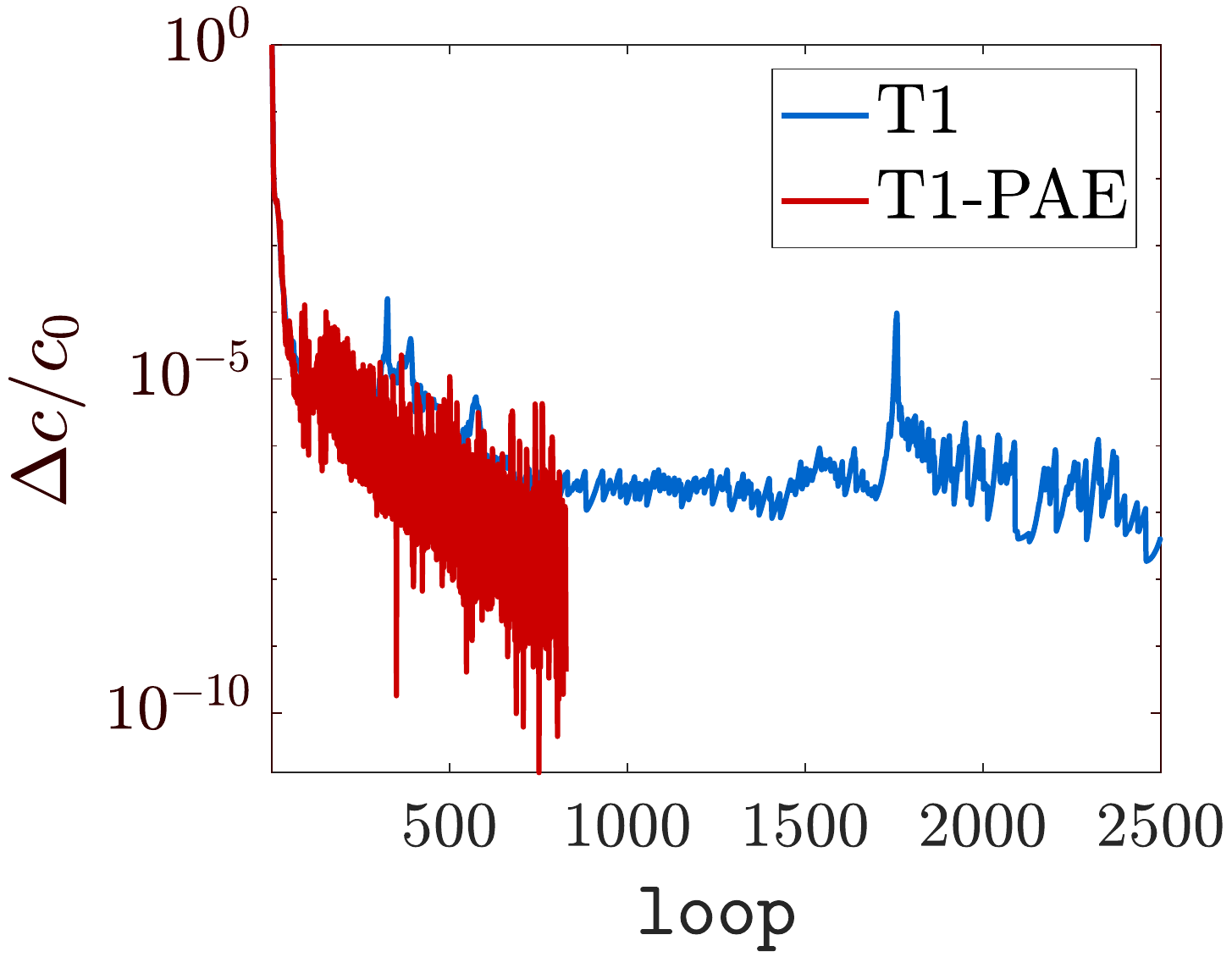}} \quad
  \subfloat[T2]{
   \includegraphics[scale = 0.25, keepaspectratio]
    {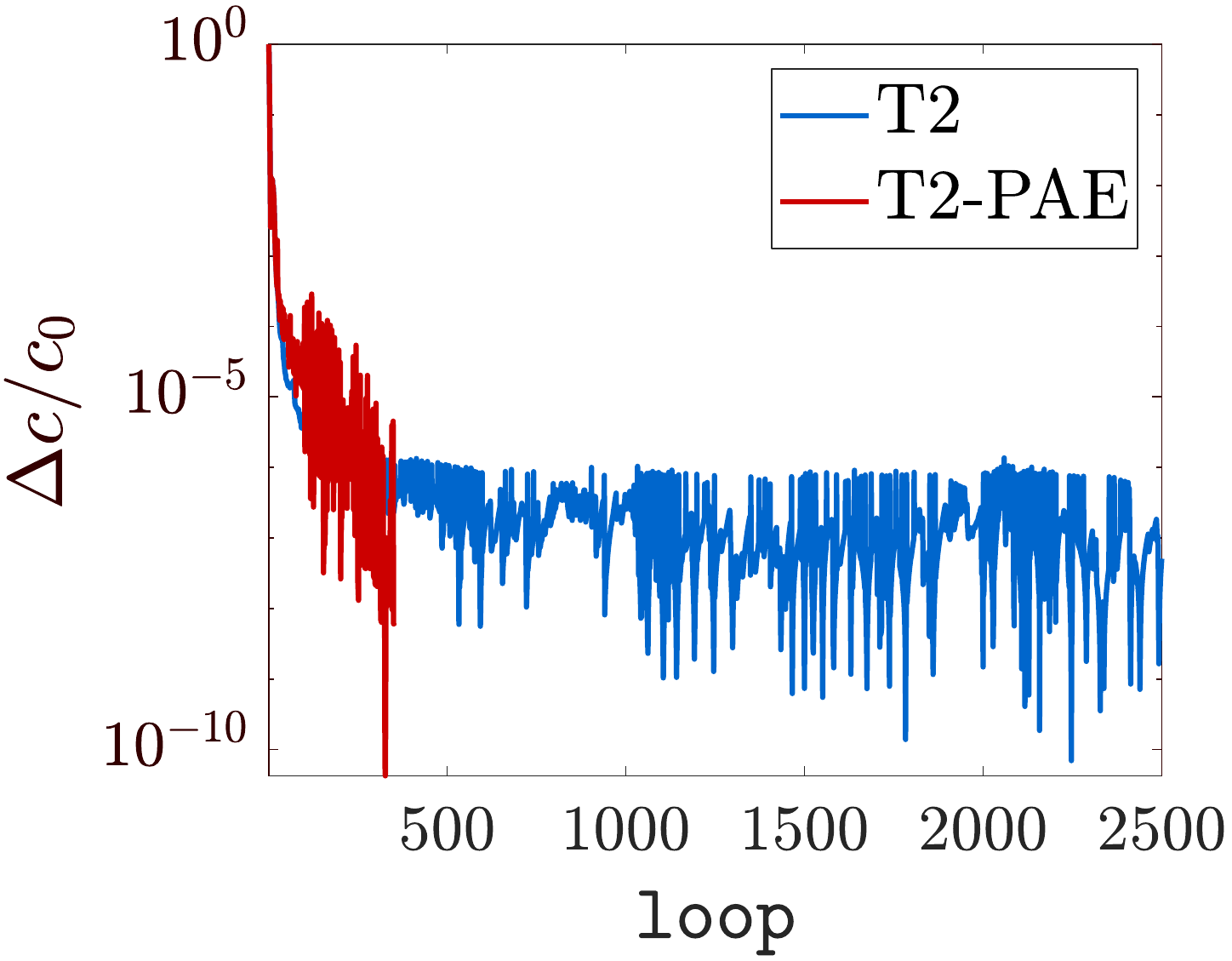}} \quad
  \subfloat[T3]{
   \includegraphics[scale = 0.25, keepaspectratio]
    {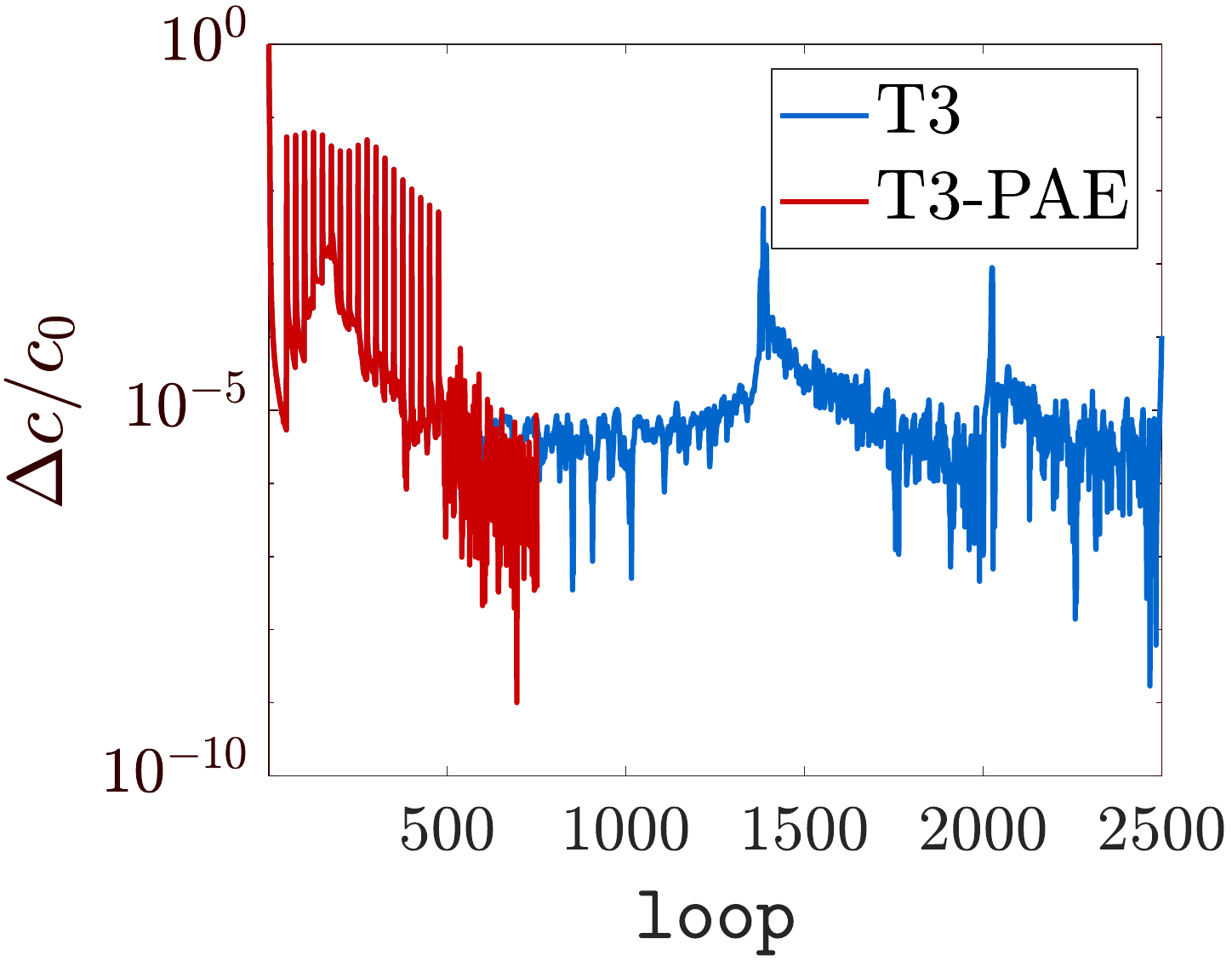}} \quad
  \subfloat[T4]{
   \includegraphics[scale = 0.25, keepaspectratio]
    {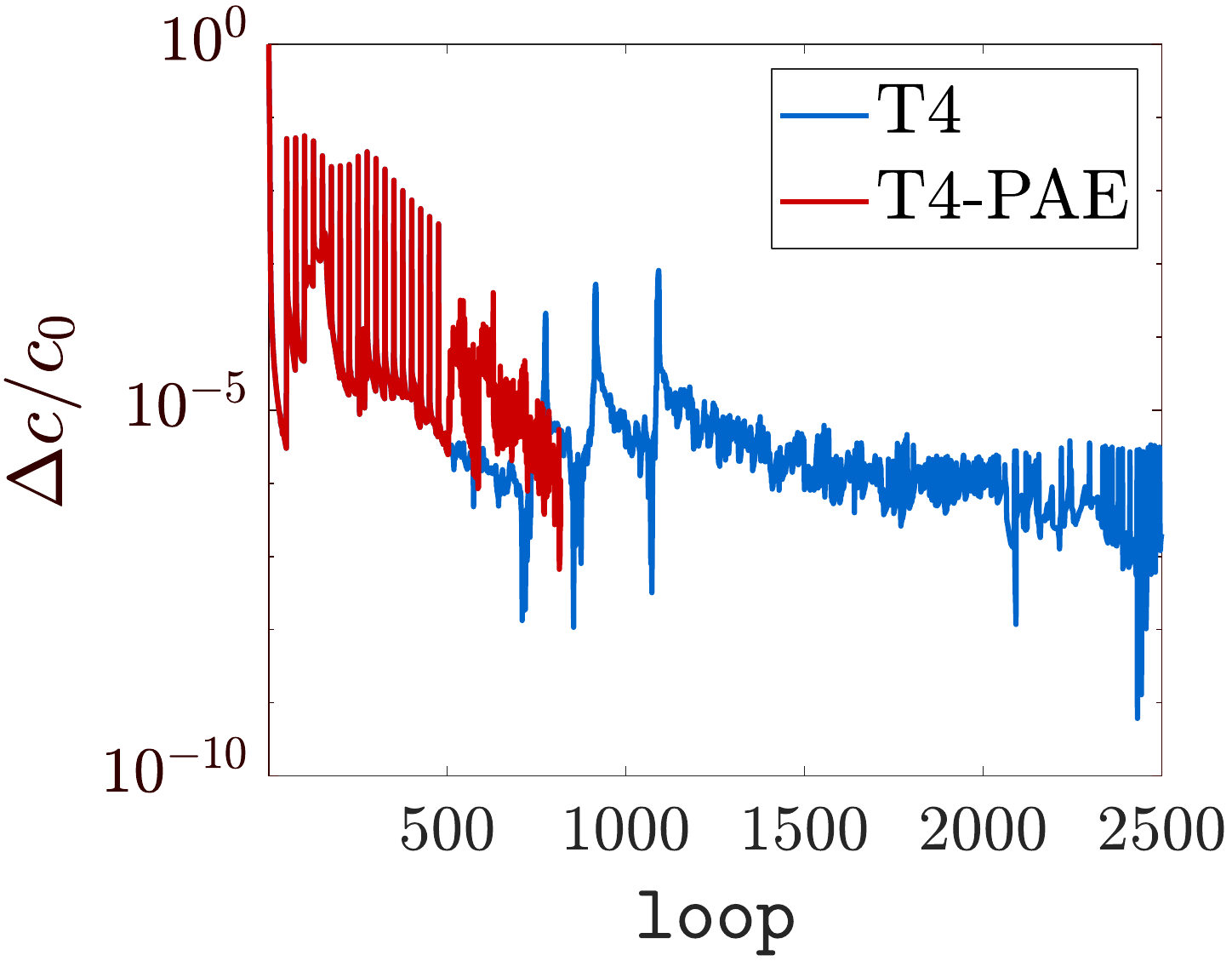}}
  \\
  \subfloat[T1]{
   \includegraphics[scale = 0.25, keepaspectratio]
    {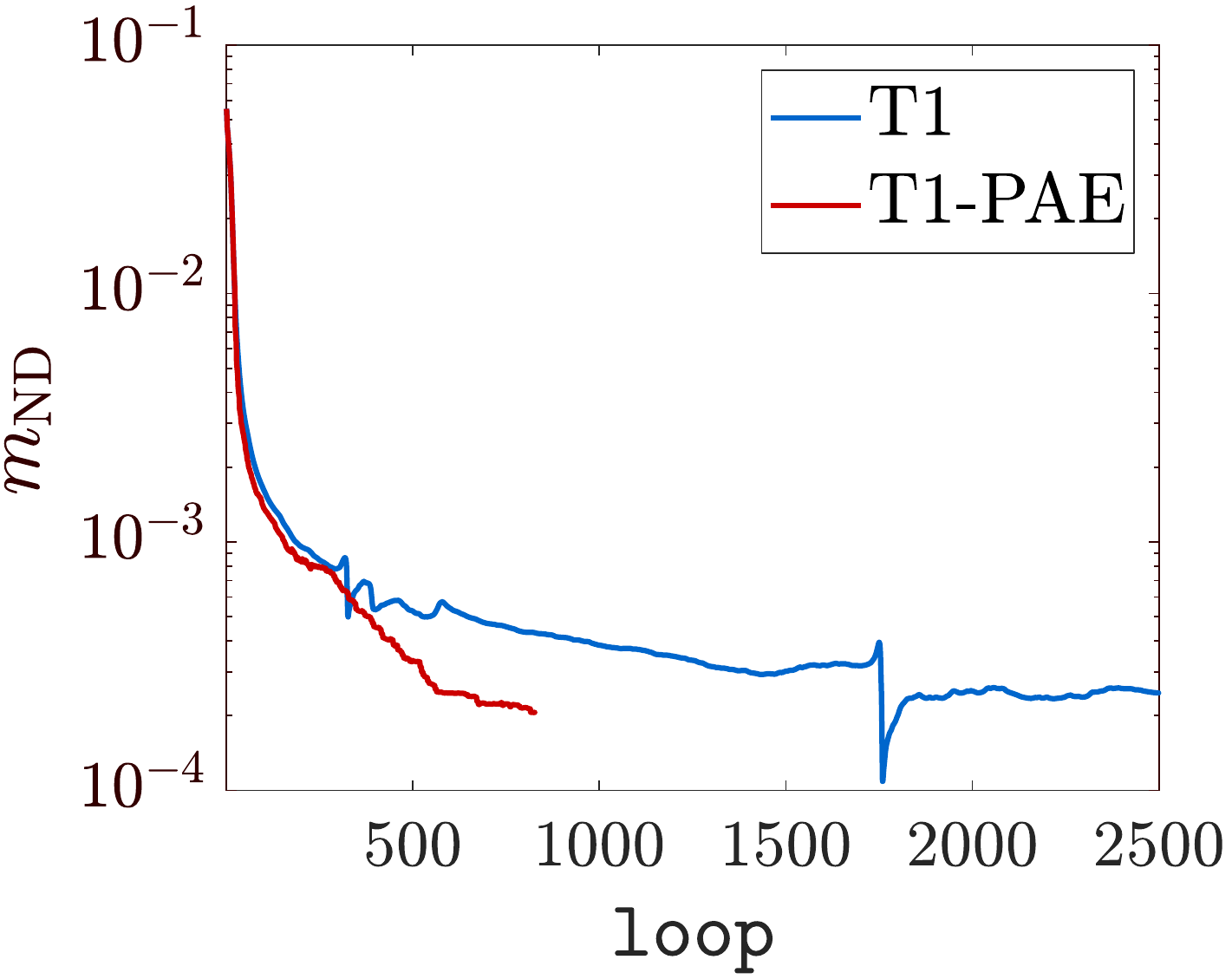}} \quad
  \subfloat[T2]{
   \includegraphics[scale = 0.25, keepaspectratio]
    {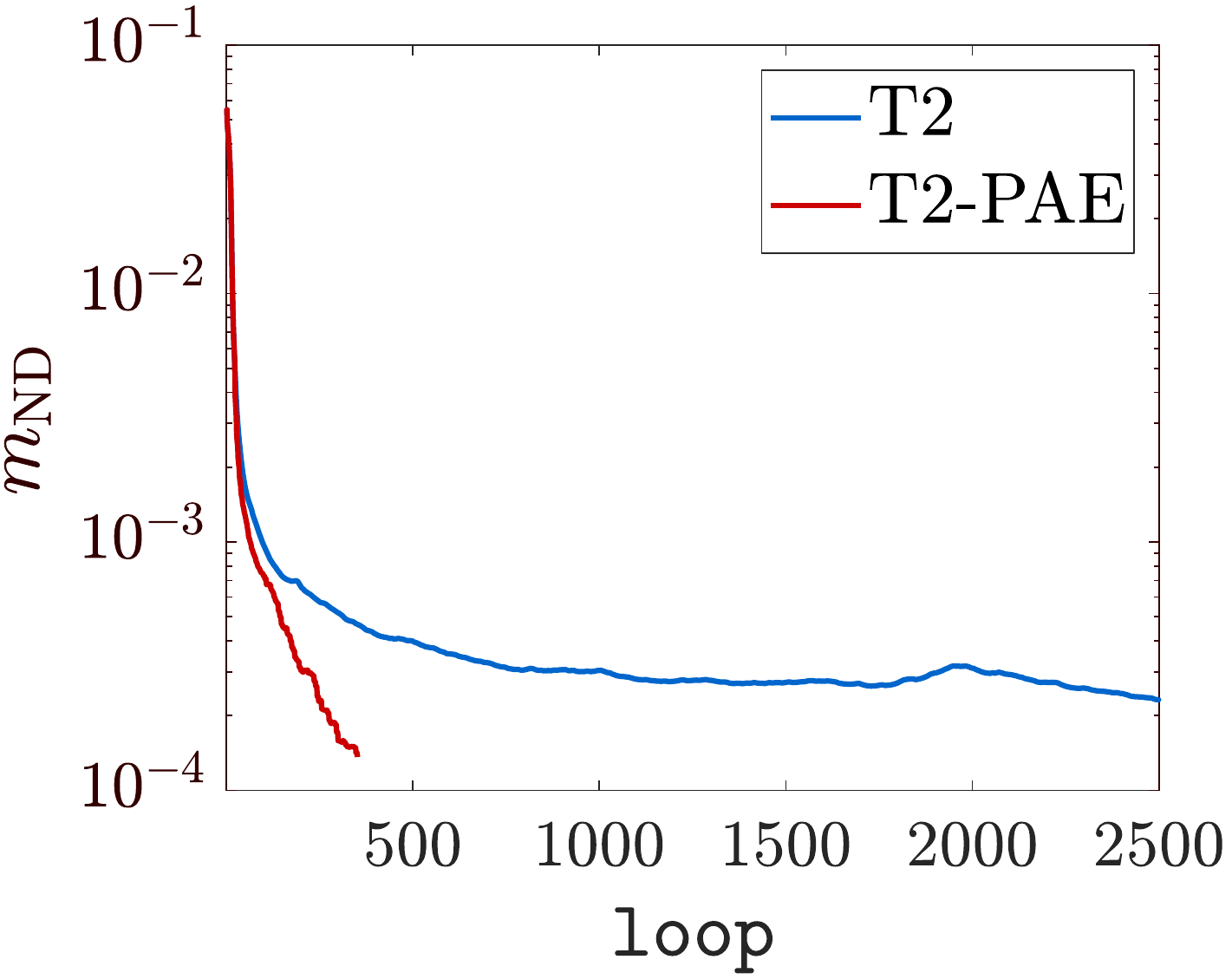}} \quad
  \subfloat[T3]{
   \includegraphics[scale = 0.25, keepaspectratio]
    {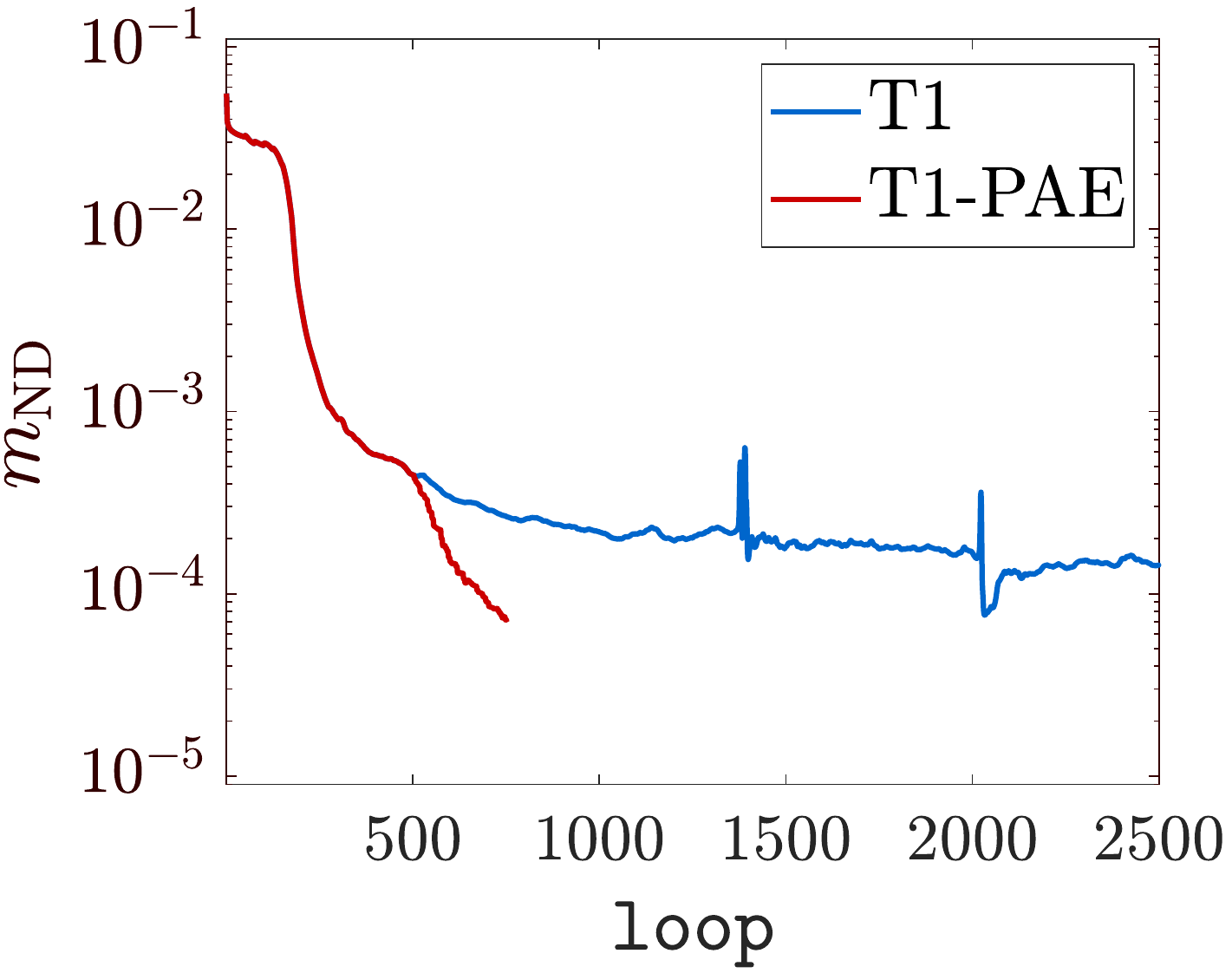}} \quad
  \subfloat[T4]{
   \includegraphics[scale = 0.25, keepaspectratio]
    {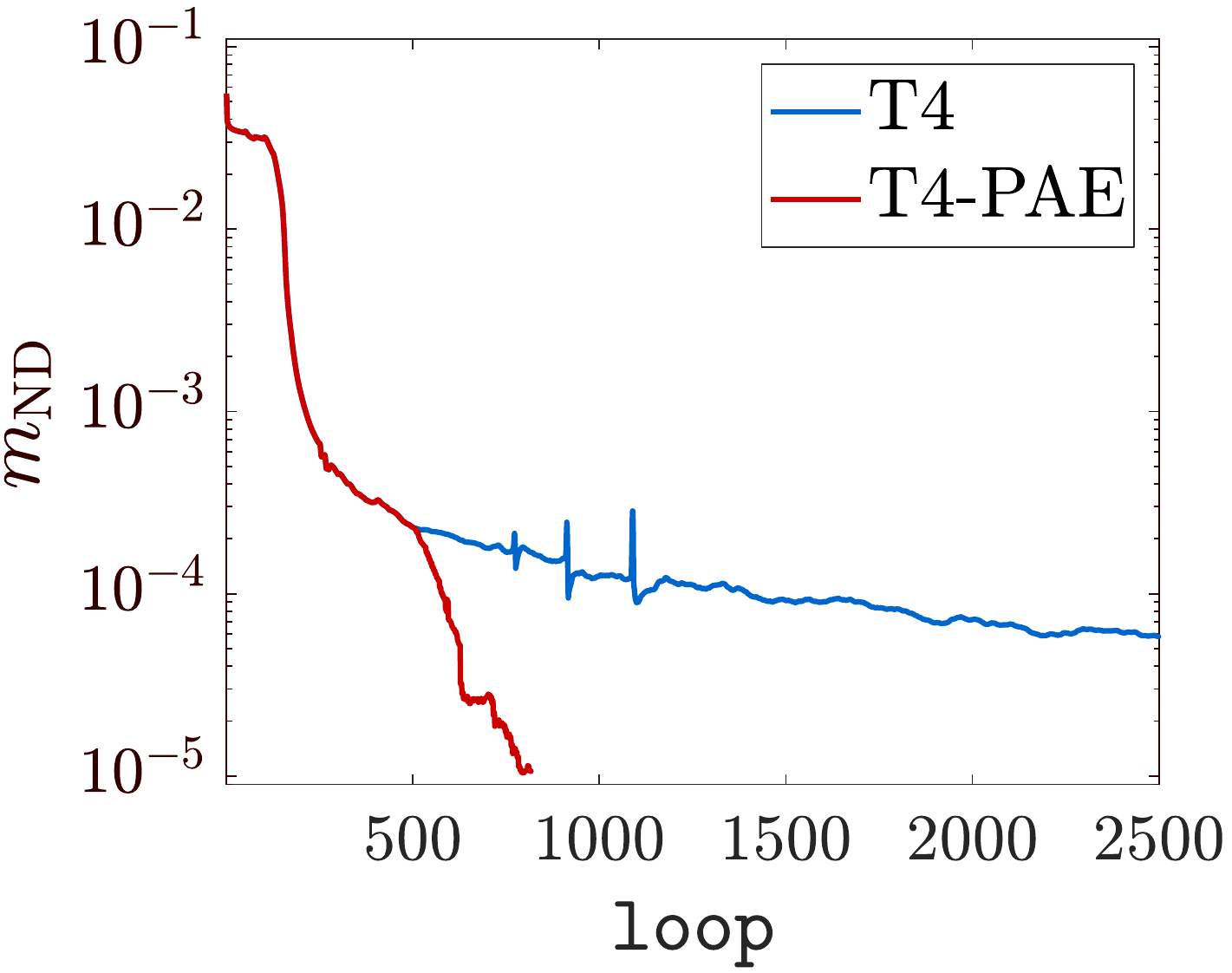}}
 \caption{\small{Evolution of some parameters related to convergence for the standard and Anderson accelerated TO process. The first row shows the normalized norm of the residual defined on physical variables, the second row shows a measure of the flatness of the objective function and the last row shows the non--discretness measure}}
 \label{fig:testPAEconvergence}
\end{figure*}

Results are collected in \autoref{tab:comparisonOCPAE} and \autoref{fig:testPAEconvergence}, showing the evolution of the norm of the residual, the flatness of the normalized compliance $\Delta c_{k} / c_{0} = (c_{k} - c_{k-1})/c_{0}$ and the non--discretness measure $m_{\rm ND}= 100 \cdot 4\mathbf{x}^{T}(1-\mathbf{x})/m$. We observe how Anderson acceleration substantially reduces the number of iterations needed to fullfill the stopping criterion, at the price of just a moderate increase in compliance ($0.2$--$3\%$). Moreover, starting the acceleration just a few iterations later (e.g. ${\rm it} = 50$ or ${\rm it} = 100$ for T1), gives much lower compliance values ($c = 254.3$ and $c = 252.9$, respectively) and for T3, T4 when the acceleration is started as the design has stabilized, compliance differences are negligible.

From \autoref{fig:testPAEtop} it is easy to notice the trend of PAE of producing a design with some more bars. This may even give slightly stiffer structures, such as for case T3, where the non accelerated approach removes some bars after ${\rm it} = 2000$, whereas stopping at the design of T3--PAE gives a stiffer structure.

A comment is about the convergence criterion used, which is different from the one in \texttt{top88} (maximum absoulute change of the design variables ($\| \mathbf{x}_{k+1} - \mathbf{x}_{k} \|_{\infty}$). Here we consider it more appropriate to check the residual with respect to the \textit{physical} design field, and the 2--norm seems to give a more global measure, less affected by local oscillations.

\subsection{Performance comparison to \texttt{top88}}
 \label{sSec:PerformanceComparison}

We compare the performance of \texttt{top99neo} to the previous \texttt{top88} code. In the following we will refer to ``\texttt{top88}'' as the original code provided by \cite{andreassen-etal_11a} and to ``\texttt{top88U}'' as its updated version making use of the \texttt{sparse2} function \citep{davis_09a} for the assembly, with \texttt{iK} and \texttt{jK} specified as integers, and the filter implemented by using \texttt{conv2}.

The codes are tested by running 100 iterations for the MBB beam example (see \autoref{fig:sketchExamples}), for the discretizations $300\times 100$, $600\times 200$ and $1200\times 400$, a volume fraction $f = 0.5$ and considering mesh independent filters of radii $r_{\rm min} = 4$, $8$ and $16$, respectively. For \texttt{top88} and \texttt{top88U} we only consider density filtering, whereas for the new \texttt{top99neo} we also consider the Heaviside projection, with the $\eta^{\ast}$ computed as described in \autoref{sSec:filtersVP}. It will be apparent that the cost of this last operation is negligible.

Timings are collected in \autoref{tab:comparisonTopcodes} where $t_{\rm it}$ is the average cost per iteration, $t_{A}$ and $t_{S}$ are the overall time spent by the assembly and solver, respectively, and $t_{U}$ is the overall time spent for updating the design variables. For \texttt{top88} and \texttt{top88U} the latter consists of the OC updating and the filtering operations performed when applying the bisection on the volume constraint. For \texttt{top99neo} this term accounts for the cost of the OC updating, that for estimating the Lagrange multiplier $\lambda^{\ast}$ as discussed in \autoref{sSec:filtersVP} and the filter and projection (Lines 59--70). $t_{P}$ collects all the preliminary operations, such as the set up of the discretization, filter etc, repeated only once, before the TO loop starts.

From $t_{\rm it}$ we clearly see that \texttt{top99neo} enhanches the performance of the original \texttt{top88} by $2.66$, $3.85$ and $5.5$ times on the three discretizations, respectively. Furthermore, timings of \texttt{top88} on the largest discretization ($1200\times 400$), relate to a smaller filter size ($r_{\rm min} = 12$), because of memory issues; thus, the speedup is even underestimated in this case. Comparing to \texttt{top88U} version, the improvements are less pronounced (i.e. $1.55$, $1.57$ and $1.78$ times) but still substantial. The computational cost of the new assembly strategy is very low, even comparing to the \texttt{top88U} version, and its weight on the overall computational cost is basically constant. Also, from \autoref{tab:comparisonTopcodes} it is clear that the design variables update weighs a lot on the overall CPU time, for both \texttt{top88} and \texttt{top88U}. On the contrary, this becomes very cheap in the new \texttt{top99neo} thanks to the strategies discussed in \autoref{sSec:filtersVP}; $t_{U}$ takes about $4$--$5\%$ of the overall CPU time. 

Computational savings would become even higher when adopting the larger filter size $r_{\rm min} = 8.75$ for the mesh $300\times 100$, and scaling to $r_{\rm min} = 17.5$ and $r_{\rm min} = 35$ on the two finer discretizations. For these cases, speedups with respect to \texttt{top88} amount to $4.45$ and $10.35$ on the first two meshes, wherease for the larger one, the setup of the filter in \texttt{top88} causes a memory overflow. Speedups with respect to \texttt{top88U} amount to $1.55$, $2.55$ and $3.6$ times respectively.

\begin{footnotesize}
\begin{table*}[tb]
 \caption{\small{Comparison of numerical performance between the old \texttt{top88}/\texttt{top88U} and new \texttt{top99neo} Matlab code. $t_{\rm it}$ is the cost per iteration, $t_{A}$, $t_{S}$, $t_{U}$ are the overall times for assembly, equilibrium equation solve and design update, respectively. $t_{P}$ is the time spent for all the preliminary operations. Values within brackets represent the $\%$ weight of the corresponding operation on the overall CPU. On the larger mesh, \texttt{top88} is run with $r_{\rm min} = 12$, because of memory issues}}
 \label{tab:comparisonTopcodes}
 \centering
  \begin{tabular}{l|ccc|ccc|ccc}
   \hline\noalign{\smallskip}
   $\Omega_{h}$ & \multicolumn{3}{c}{$300 \times100$, $r_{\rm min} = 4$} & \multicolumn{3}{c}{$600 \times200$, $r_{\rm min} = 8$} & \multicolumn{3}{c}{$1200 \times400$, $r_{\rm min} = 16$} \\
   \noalign{\smallskip}\hline
                & \texttt{top88} & \texttt{top88U} & \texttt{top99neo} & \texttt{top88} & \texttt{top88U} & \texttt{top99neo} &
                \texttt{top88} & \texttt{top88U} & \texttt{top99neo} \\
   \noalign{\smallskip}\hline
   $t_{\rm it}$ & 0.615  & 0.358      & 0.231      & 4.57        & 1.87        & 1.19        & 31.3        & 10.1        & 5.69        \\
   $t_{A}$  & 19.4(31.5) &  5.4(15.0) &  1.4 (6.1) &  83.1(18.2) &  31.3(16.7) &   5.6 (4.7) & 361.1(11.6) & 151.5(15.2) &  30.7 (5.4) \\
   $t_{S}$  & 23.1(37.4) & 22.9(59.3) & 19.7(85.3) & 122.4(26.8) & 109.3(58.4) & 106.9(89.7) & 592.5(19.0) & 513.2(50.9) & 510.5(89.6) \\
   $t_{U}$  & 13.3(21.6) &  4.8(13.5) &  1.2 (4.8) & 223.8(48.8) &  38.0(20.3) &   5.2 (4.4) &1164.2(37.4) & 310.4(31.4) &  29.2 (5.1) \\
   $t_{P}$  &  0.8(1.3)  & 0.06 (0.2) &  0.1 (0.3) &  12.9 (2.8) & 0.1($<0.1$) & 0.2($<0.1$) &  92.3 (3.1) & 0.5($<0.1$) & 0.6($<0.1$) \\
   \noalign{\smallskip}\hline
  \end{tabular}
\end{table*}
\end{footnotesize}

\subsection{Frame reinforcement problem}
 \label{sSec:FrameReinforcement}

Let us go back to the example of \autoref{fig:settingFigures}(a), adding the specification of passive domains and a different loading condition.

We may think of a practical application like a reinforcement problem for the solid frame, with thickness $t=L/50$ ($\mathcal{P}_{1}$), subjected to two simultaneous loads. A vertical, uniformly distributed load with density $q = -2$ and a horizontal height--proportional load, with density $b = \pm y/L$. Some structural material has to be optimally placed within the active design domain $\mathcal{A}$ in order to minimize the compliance, while keeping the void space ($\mathcal{P}_{0}$), which may represent a service opening.

To describe this configuration we only need to replace Lines 31--33 with the following
\begin{lstlisting}[basicstyle=\scriptsize\ttfamily,breaklines=true,numbers=none,frame=single]
elNrs = reshape(1:nEl,nely,nelx);
[lDofv,lDofh]=deal(2*nodeNrs(1,:),2*nodeNrs(:,end)-1);
fixed = [1,2,nDof];
a1=elNrs(1:nely/50,:);
a2=elNrs( :,[1:nelx/50,end-nelx/50+1:end]);
a3=elNrs(2*nely/5:end,2*nelx/5:end-nelx/5);
[pasS,pasV]=deal(unique([a1(:);a2(:)]),a3(:));
\end{lstlisting}
where \texttt{lDofv} and \texttt{lDofh} target the DOFs subjected to vertical and horizontal forces, respectively. Then, the load (Line 34) is replaced with
\begin{lstlisting}[basicstyle=\scriptsize\ttfamily,breaklines=true,numbers=none,frame=single]
F=fsparse(lDofv',1,-2/(nelx+1),[nDof,1])+fsparse(...
lDofh,1,-[0:1/(nely).^2:1/nely]',[nDof,1]);
\end{lstlisting}

\begin{figure}[t]
 \centering
  \subfloat[]{
   \includegraphics[scale = 0.25, keepaspectratio]
   {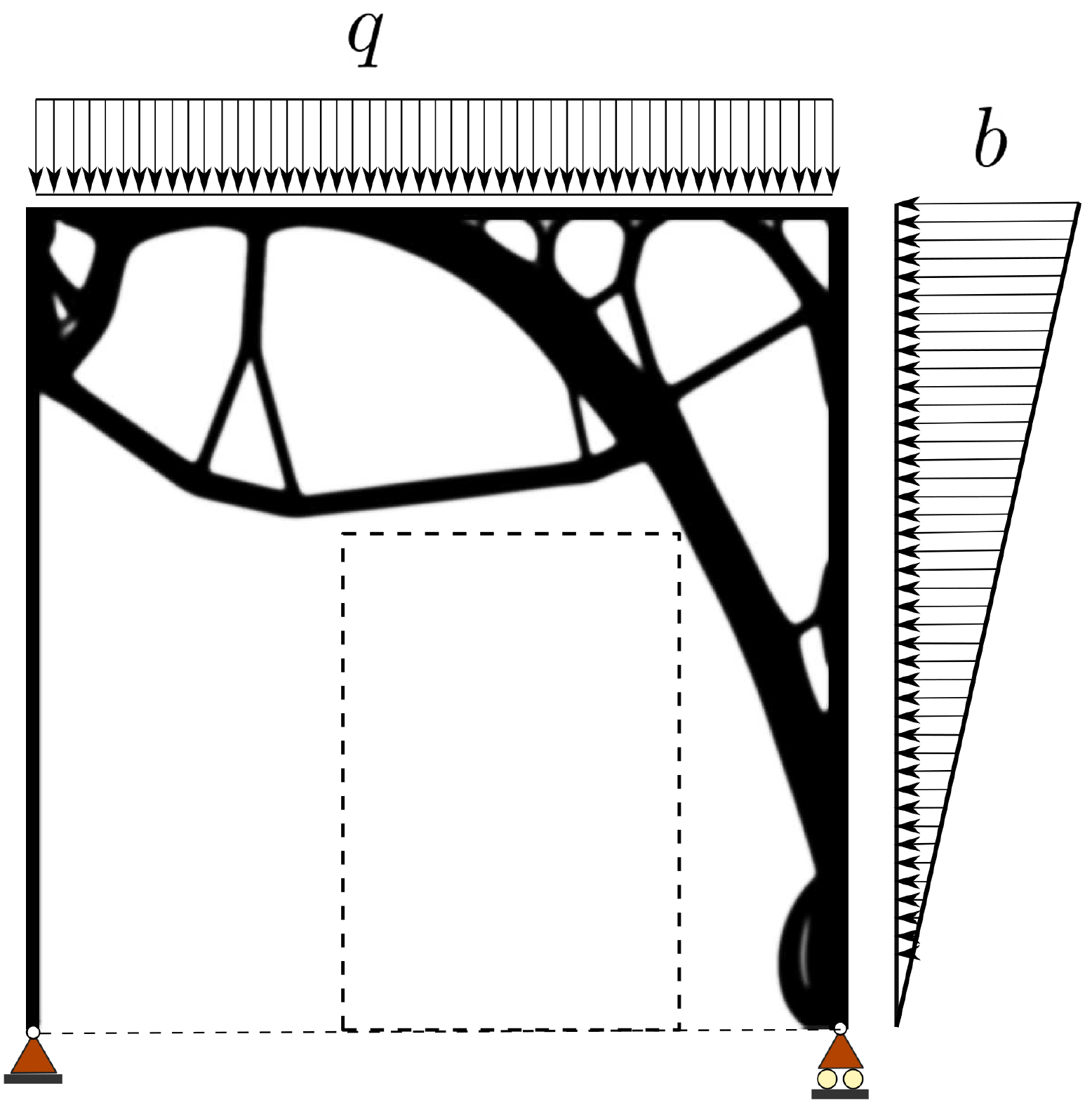}}
  \subfloat[]{
   \includegraphics[scale = 0.25, keepaspectratio]
   {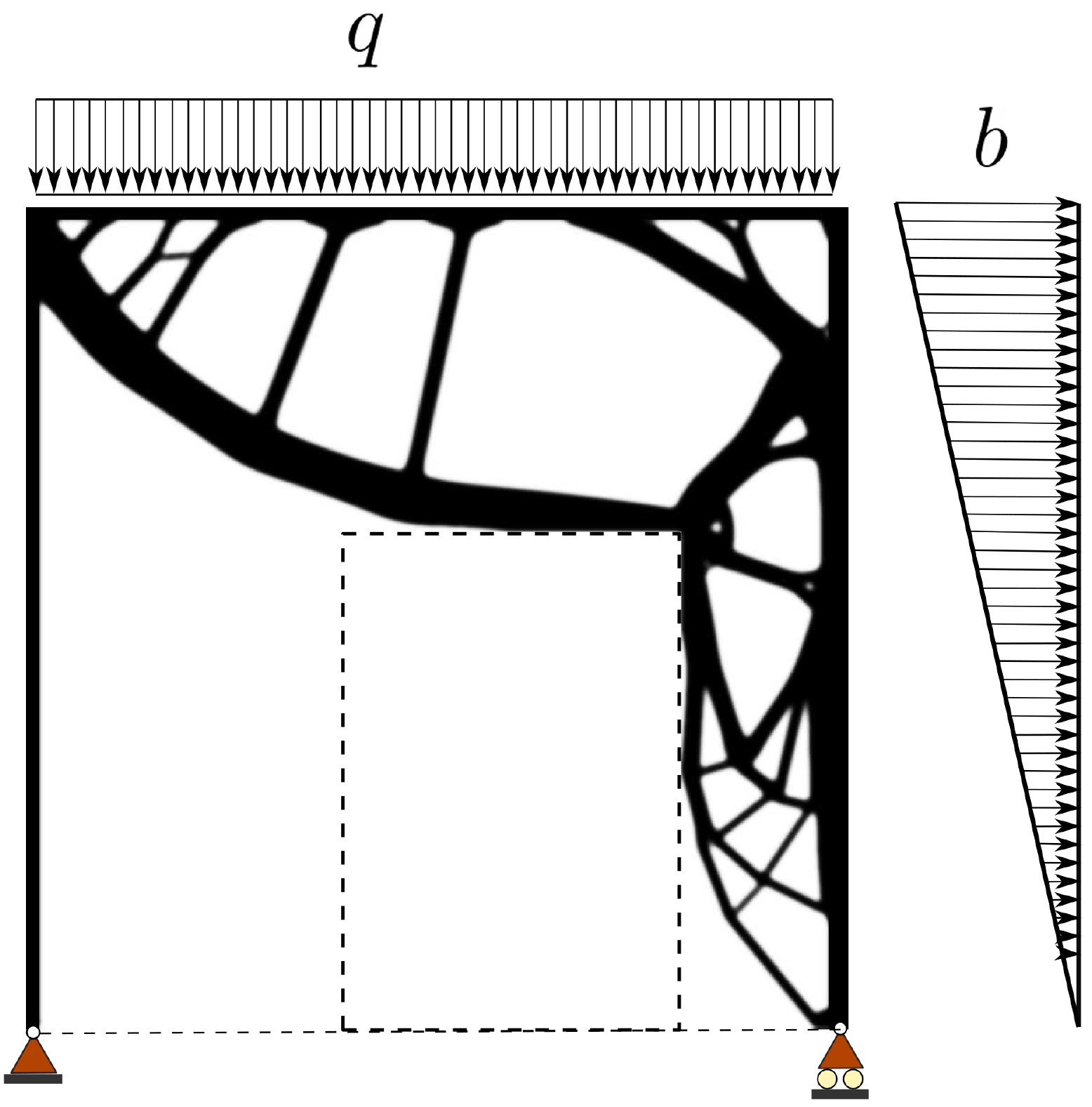}}
 \caption{\small{Designs obtained for the frame reinforcement problem sketched in \autoref{fig:settingFigures}(a)}. In (a) the horizontal, triangular load distribution is pointing leftwards, wherease in (b) it s pointing rightwards}
 \label{fig:frameReinforcement}
\end{figure}

\autoref{fig:frameReinforcement} shows the two optimized design corresponding to the two orientations of the horizontal load $b$, after 100 re--design steps. The routine \texttt{top99neo} has been called with the following arguments \texttt{nely=nelx=900}, \texttt{volfrac=0.2}, \texttt{penal=3}, \texttt{rmin=8}, \texttt{ft=3}, \texttt{eta=0.5}, \texttt{beta=2} and no continuation is applied. The cost per iteration is about $10.8$s and, considering the fairly large discretization of $1.62\cdot 10^{6}$ DOFs, is very reasonable.

\section{Extension to 3D}
 \label{Sec:3Dextension}

The implementation described in \autoref{Sec:codeStructure} is remarkably easy to extended to 3D problems (see \autoref{Sec:3DcodeComplianceMinimization}).

Notable modifications are the definition of $K^{(s)}_{e}$ for the 8--node hexahedron (Lines 24--47) and the solution of the equilibrium equations \eqref{eq:stateProblem}, now performed by
\begin{lstlisting}[basicstyle=\scriptsize\ttfamily,breaklines=true,numbers=none,frame=single]
L = chol( K( free, free ), 'lower' );
U( free ) = L' \ ( L \ F( free ) );
\end{lstlisting}
which in this context has been observed to be faster than the \texttt{decomposition} routine. Then, apart from the plotting instructions, all the operations are the same as in the 2D code and only 12 lines need minor modifications, basically to account for the extra space dimension (see tags ``$\texttt{\#3D\#}$'' in \autoref{Sec:3DcodeComplianceMinimization}).

\begin{figure*}[t]
 \centering
  \subfloat[]{\fbox{
   \includegraphics[scale = 0.175, keepaspectratio]
   {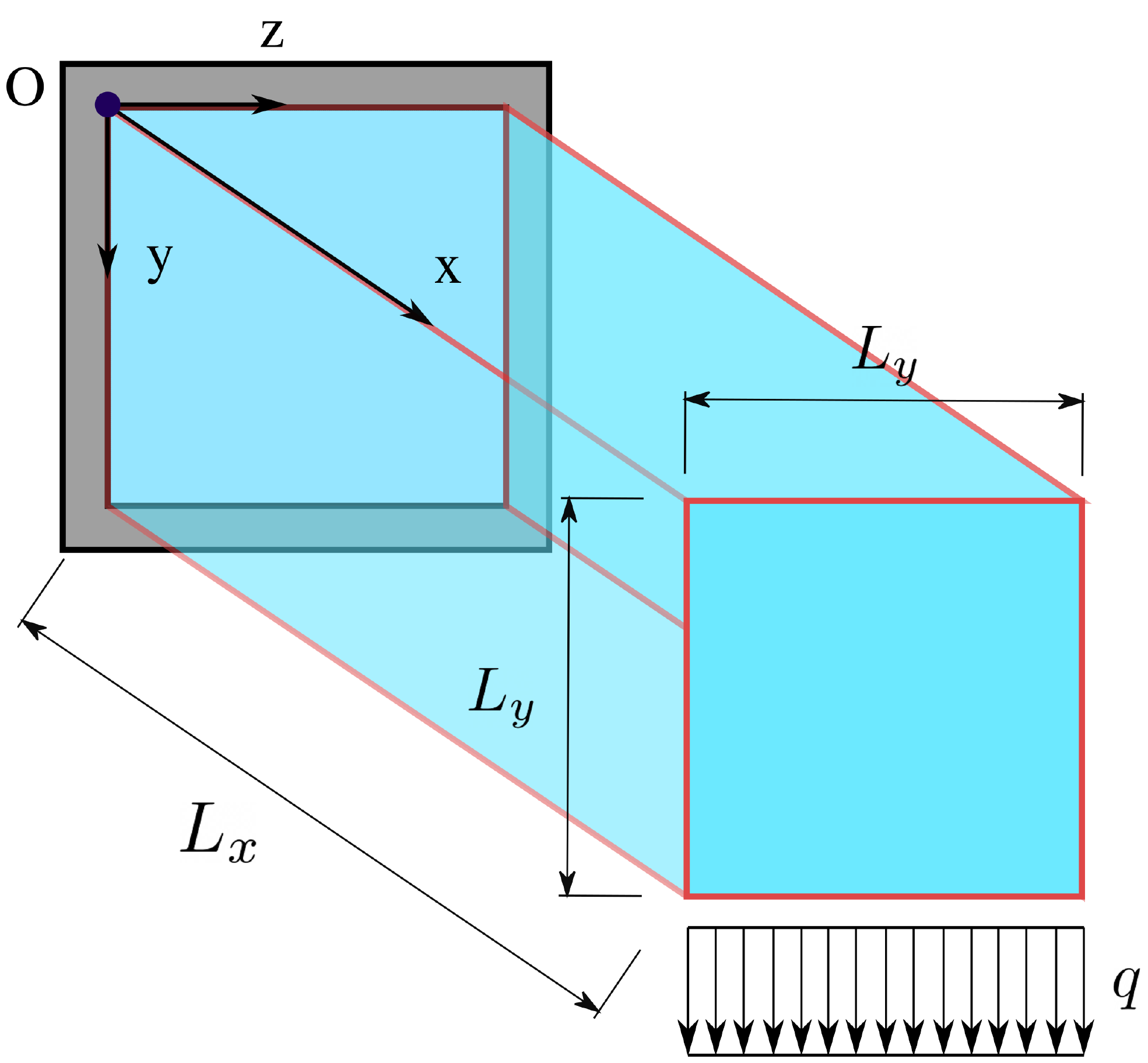}}} \quad
  \subfloat[Neumann ($c=3993.7$)]{
   \includegraphics[scale = 0.375, keepaspectratio]
   {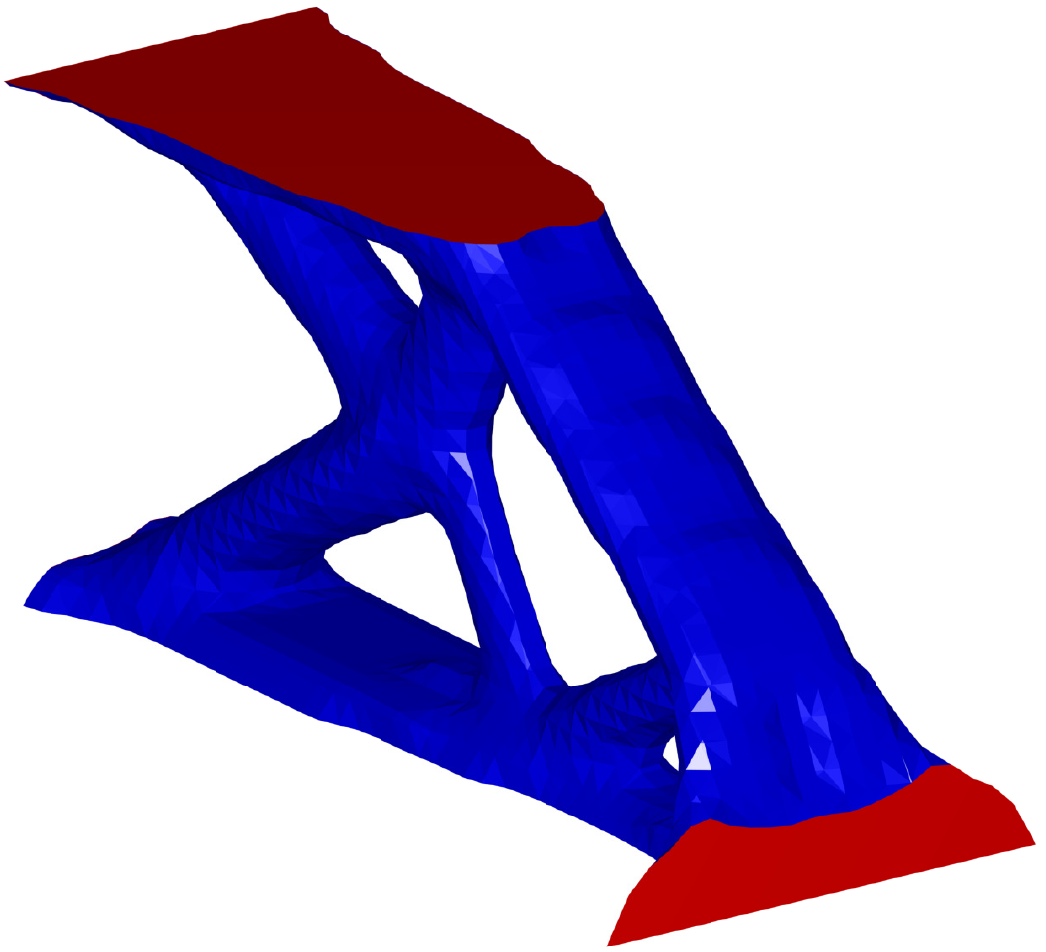}} \quad
  \subfloat[Dirichlect ($c=4074.3$)]{
   \includegraphics[scale = 0.375, keepaspectratio]
   {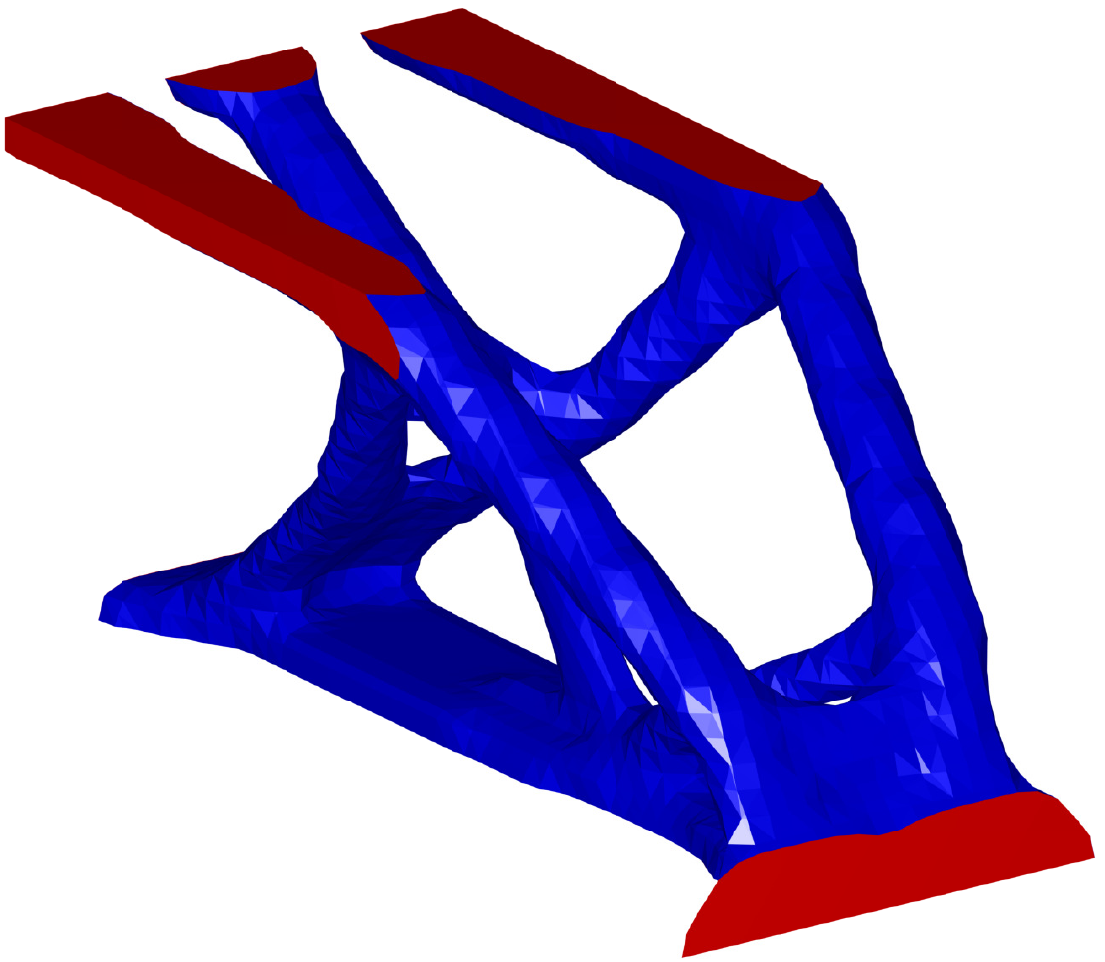}} \quad
  \subfloat[Neumann]{
   \includegraphics[scale = 0.375, keepaspectratio]
   {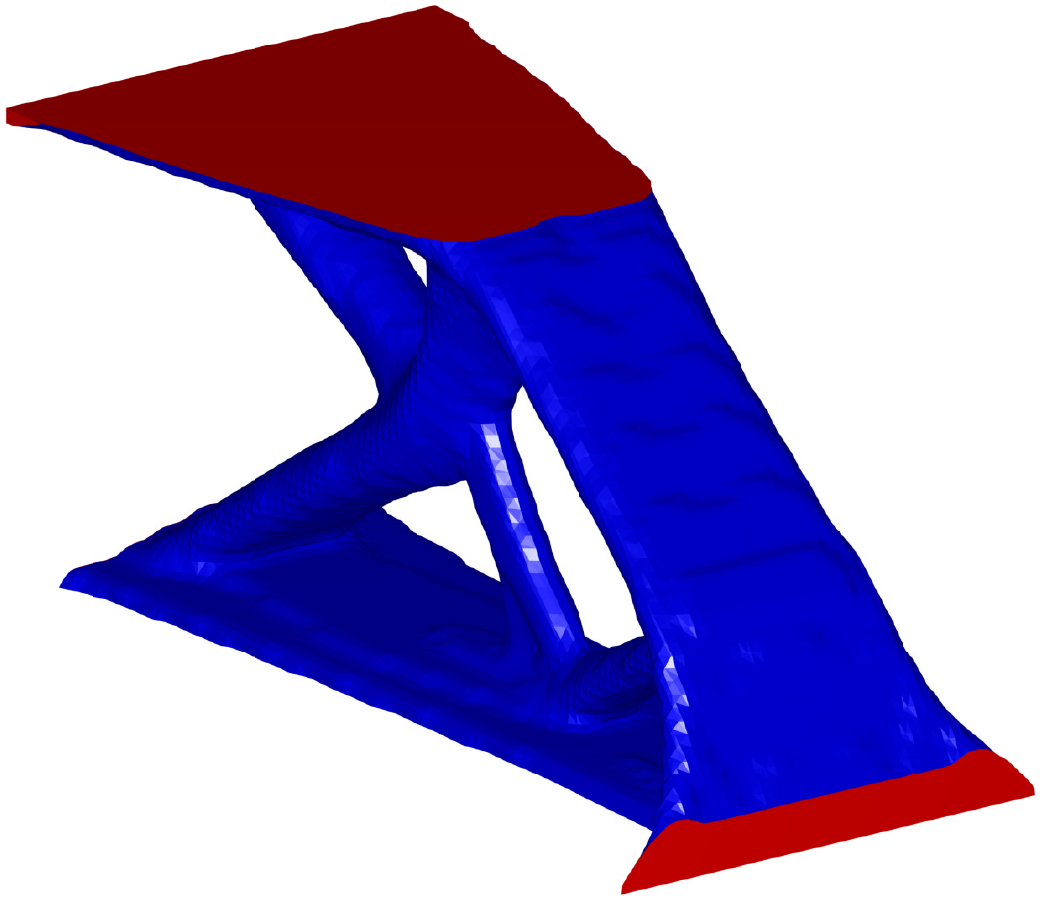}}
 \caption{\small{Geometrical sketch of the 3D cantilever example (a) and optimized topology for $\Omega_{h} = 48\times 24\times 24$ and considering the two filter boundary conditions (b,c). The design in (d) corresponds to the finer mesh $\Omega_{h} = 96\times 48\times 48$ and has been obtained by replacing the direct solver with the multigrid--preconditioned CG (see \cite{amir-etal_14a} for details)}}
 \label{fig:cantilever3D}
\end{figure*}

We test the 3D implementation on the cantilever example shown in \autoref{fig:cantilever3D}(a), for the same data considered in \citep{amir-etal_14a}. The discretization is set to $\Omega_{h} = 48\times 24\times 24$, the volume fraction is $f = 0.12$ and the filter radius $r_{\rm min} = \sqrt{3}$. We also consider the volume--preserving Heaviside projection, (\texttt{ft=3}). \autoref{fig:cantilever3D} (b,c) show the designs obtained after 100 redesign steps, for the two different filter boundary conditions. The design in (b), identical to the one in \citep{amir-etal_14a}, corresponds to zero--Neumann boundary conditions (i.e. the option ``\texttt{symmetric}'' was used in \texttt{imfilter}). The design in (c) on the other hand, corresponds to zero--Dirichlect boundary conditions for the filter operator and is clearly a worse local minimum.

The overall CPU time spent over 100 iterations is $1741$s and about $96\%$ of this is due to the solution of the state equation. Only $1.2\%$ of the CPU time is taken by matrix assemblies and $0.4\%$ by filtering and the design update processes.

Upon replacing the direct solver in \texttt{top3D125} with the same multigrid preconditioned CG solver of \cite{amir-etal_14a} we can compare the efficiency of the two codes. We refer to \autoref{tab:comparisonTopcodes3D} for the CPU timings, considering the discretizations $\Omega_{h} = 48\times 24\times 24$ ($l=3$ multigrid levels) and $\Omega_{h} = 96 \times 48 \times 48$ ($l=4$ multigrid levels). \texttt{top3D125} shows speedups of about $1.8$ and $1.9$, respectively and most of the time is cut on the matrix assembly. In the code of \cite{amir-etal_14a} this operation takes about $50\%$ of the overall time (and notably has the same weight as the state equation solve) whereas in \texttt{top3D125} this weight is cut to $7-10\%$. Also the time spent for the OC update is reduced, even though the code of \cite{amir-etal_14a} already implemented a strategy for avoiding filtering at each bisection step.

\begin{footnotesize}
\begin{table}[tb]
 \caption{\small{Performance comparison between the new \texttt{top3D125} code and the one from \cite{amir-etal_14a}. $t_{\rm it}$, $t_{A}$, $t_{S}$, $t_{U}$ and $t_{P}$ have the same meaning as in \autoref{tab:comparisonTopcodes} and numbers between brackets denote the $\%$ weight of the operations on the overall CPU time}}
 \label{tab:comparisonTopcodes3D}
 \centering
  \begin{tabular}{l|cc|cc}
   \hline\noalign{\smallskip}
   $\Omega_{h}$ & \multicolumn{2}{c}{$48 \times 24 \times 24$, $r_{\rm min} = \sqrt{3}$} & \multicolumn{2}{c}{$96 \times 48 \times 48$, $r_{\rm min} = 2\sqrt{3}$} \\
   \noalign{\smallskip}\hline
                & \texttt{top3dmgcg} & \texttt{top3D125} & \texttt{top3dmgcg} & \texttt{top3D125} \\
   \noalign{\smallskip}\hline
   $t_{\rm it}$ &     3.19    & 1.79        & 27.33      & 14.20 \\
   $t_{A}$      & 160.6(50.3) &  13.1 (7.4) & 1369(50.1) & 137.2(9.7)  \\
   $t_{S}$      & 148.1(46.4) & 151.7(84.7) & 1250(45.7) &  1272(89.5) \\
   $t_{U}$      &  1.97 (0.6) &   0.7 (0.4) & 21.2 (0.8) &  15.12(1.1) \\
   $t_{P}$      &  0.74 (0.4) &  0.24 (0.1) & 39.2 (1.4) &  0.29($<$0.1) \\
   \noalign{\smallskip}\hline
  \end{tabular}
\end{table}
\end{footnotesize}

\section{Concluding remarks}
 \label{Sec:conclusions}

We have presented new Matlab implementations of compliance Topology Optimization for 2D and 3D domains. Compared to the previous \texttt{top88} code \citep{andreassen-etal_11a} and available 3D codes (e.g. by \cite{liu-tovar_14a} or \cite{amir-etal_14a}), the new codes show remarkable speedups.

Improvements are mainly due to the following
\begin{enumerate}
 \item The matrix assembly is made much more efficient by defining mesh related quantities as integers (Matlab \texttt{int32}) and assembling just one half of the matrix;
 \item The number of OC iterations is drastically cut by looking at the explicit expression of the Lagrange multiplier for the problem at hand;
 \item Filter implementation and volume-preserving density projection allow to speed up the redesign step.
\end{enumerate}

The new codes are computationally well balanced and as the problem size increases the majority of the time ($85$ to $90\%$ for 2D and even $96\%$ for 3D discretizations) is spent on the solution of the equilibrium system. This is precisely what we aimed at, as this step can be dealt with efficiently by preconditioned iterative solvers \citep{amir-etal_14a,ferrari-etal_18a,ferrari-sigmund_20a}. We also discussed Anderson acceleration, that has recently been applied to TO also by \cite{thesis:li-paulino_18}, to accelerate the convergence of the overall optimization loop.

We point out that even if we specifically addressed volume constrained compliance minimization and density-based TO the methods above can be applied also to level-set and other TO approaches. Point 1 can be extended to all problems governed by symmetric matrices. Point 2 and 3 can also be extended to other problems, to some extent and Anderson acceleration is also usable in a more general setting (e.g. within MMA).

Therefore, we believe that this contribution should be helpful to all researchers and practitioners who aim at tackling TO problems on laptops, and set a solid framework for the efficient implementation of more advanced procedures.

\begin{small}
\paragraph{\textsc{\textbf{Reproducibility of results}}}
Matlab codes are listed in the Appendix and available at \texttt{www.topopt.dtu.dk}. The \texttt{stenglib} package, containing the \texttt{fsparse} function, is avaialble for download at \texttt{https://github.com/stefanengblom/stenglib}.
\end{small}

\begin{small}
\paragraph{\textsc{\textbf{Conflict of interest}}}
On behalf of all authors, the corresponding author states that there is no conflict of interest.
\end{small}

\begin{acknowledgements}
 The project is supported by the Villum Fonden through the Villum Investigator Project ``InnoTop''. The authors are grateful to members of the TopOpt group for their useful testing of the code.
\end{acknowledgements}

\begin{small}
\bibliographystyle{spbasic}
\bibliography{/home/fedeferro/Dropbox/LiteratureDatabase/Database.bib}
\end{small}
\appendix

\section{Elaboration on the OC update}
 \label{Sec:lmEstimate}

Let us consider \eqref{eq:OptimizationProblem} at a given design point $\mathbf{x}_{k}$ assuming the reciprocal and linear approximation for the compliance and volume functions, respectively \citep{book:christensen-klarbring_08}
\begin{equation}
 \label{eq:app-linOptProblem}
 \begin{cases}
       & \min\limits_{\mathbf{x}\in [\delta_{-}, \delta_{+}]^{m}}
       c\left( \mathbf{x} \right) \simeq c_{k} + 
       \sum^{m}_{e=1} (-x^{2}_{k, e}\partial_{e} c(\mathbf{x}_{k})) x^{-1}_{e} \\
  {\rm s.t.} & \sum^{m}_{e=1} \partial_{e} V(\mathbf{x}_{k}) x_{e} - f|\Omega_{h}| \leq 0
 \end{cases}
\end{equation}

We set up the Lagrangian associated with \eqref{eq:app-linOptProblem}
\[
 L( \mathbf{x}, \lambda ) = c(\mathbf{x}) + \lambda \left( \sum^{m}_{e=1} \partial_{e} V(\mathbf{x}_{k}) x_{e} - f|\Omega_{h}| \right)
\]
and seek the pair $(\mathbf{x}_{k+1},\lambda^{\ast}_{k})\in\mathbb{R}^{m}\times\mathbb{R}_{+}$ solving the subproblem
\begin{equation}
 \label{eq:app-minmax}
  \max_{\lambda > 0} \left\{ \psi(\lambda) :=
  \min_{\mathbf{x}\in \mathcal{C}} L( \mathbf{x}, \lambda )\right\}
\end{equation}
where $\mathcal{C} = \{ \mathbf{x} \in \mathbb{R}^{m} \mid \delta_{-}\leq x_{e} \leq \delta_{+}, \: e = 1 \ldots, m \}$ and $\psi(\lambda)$ is the dual function. \eqref{eq:app-minmax} is solved by Primal-Dual (PD) iterations, as $\mathbf{x}$ and $\lambda$ are interlaced. Replacing $\boldsymbol{\xi} = \mathbf{x}_{k}$ and using subscripts $(j)$ to denote inner PD iterations we have

\begin{enumerate}
 \item Fixed $\lambda = \lambda_{(j)}$, the inner minimization in \eqref{eq:app-minmax} gives 
  \begin{equation*}
   \xi^{2}_{e}\partial_{e}c(\boldsymbol{\xi})x^{-2}_{e} + \lambda
   \partial_{e}V(\boldsymbol{\xi}) = 0
   \Longrightarrow
   x_{e} = \xi_{e}
   \left( -\frac{\partial_{e}c(\boldsymbol{\xi})}
   {\lambda\partial_{e}V(\boldsymbol{\xi})} \right)
   ^{\frac{1}{2}}
 \end{equation*}
 due to separability of the approximation. Let us denote the rightmost expression $x_{e} = \mathcal{F}_{(j)e}(\lambda)$, and taking into account the box constraints in $\mathcal{C}$ we have
 \begin{small}
 \begin{equation}
  \label{eq:app-primalMap}
  \mathcal{U}(x_{e}) = 
   \begin{cases}
     x_{(j+1),e} = \delta_{-} & {\rm if} \: e \in 
     \mathcal{L} = \{e\mid x_{(j+1),e} \leq \delta_{-} \} \\
     x_{(j+1),e} = \delta_{+} & {\rm if} \: e \in 
     \mathcal{U} = \{e\mid x_{(j+1),e} \geq \delta_{+} \} \\
     x_{(j+1),e} = \mathcal{F}_{(j),e}
     & {\rm if} \: e \in \mathcal{M} = \{e\mid\delta_{-} < x_{(j+1),e} < \delta_{+} \} \\
    \end{cases}
 \end{equation}
 \end{small}
 where $\mathcal{C} = \mathcal{L} + \mathcal{U} + \mathcal{M}$. The above is equivalent to \eqref{eq:OCupdate}.
 \item We then evaluate the dual function for $x_{(j+1)}$ given by \eqref{eq:app-primalMap}, and the stationarity ($\partial_{\lambda} \psi = 0$) gives
 \begin{equation*}
  \label{eq:app-stationarityDual}
 \sum^{m}_{e=1} \partial_{e} V(\boldsymbol{\xi})
 (\chi_{\mathcal{U}}\delta_{+} + \chi_{\mathcal{L}}\delta_{-} + 
 \mathcal{F}_{(j),e}(\lambda)\chi_{\mathcal{M}} ) - f|\Omega_{h}| = 0
 \end{equation*}
where $\chi_{[\cdot]}$ is the characteristic function of a set. In this simple case, the above can be solved for $\lambda_{(j+1)}$, the Lagrange multiplier enforcing the volume constraint for the updated density $x_{(j+1)}$, and after some simplifications we obtain
 \begin{equation}
  \label{eq:app-dualMap}
  \lambda_{(j+1)} = \left(
  \frac{\sum_{e\in\mathcal{M}}x_{(j+1)e}(\partial_{e}c(\boldsymbol{\xi})/\partial_{e}V(\boldsymbol{\xi}))^{1/2}}
  {f|\Omega_{h}|/\partial_{e}V(\boldsymbol{\xi}) - 
  |\mathcal{L}|\delta_{-} - |\mathcal{U}|\delta_{+}}\right)^{2}
 \end{equation}
 where $|\cdot|$ denotes the number of elements in a set.
\end{enumerate}

Equations \eqref{eq:app-primalMap} and \eqref{eq:app-dualMap} can be iteratively used to compute the new solution $(\mathbf{x}_{k+1},\lambda^{\ast}_{k})$, as implemented in the code here below (again, note that \texttt{lm} here represents $\sqrt{\lambda}$)
\begin{lstlisting}[basicstyle=\scriptsize\ttfamily,breaklines=true,numbers=none,frame=single]
u=min(xT+move,1); l=max(xT-move,0);
ocP=@(s) xT(s).*sqrt(-dc(s)./dV0(s));
lm=mean(ocP(act))/volfrac;
lmOld = 0;
while abs(lm-lmOld)>1e-10
  tmp=ocP(act)/lm;
  [setu,setl]=deal(find(tmp>u),find(tmp<l));
  setM=not(abs(sign(sign(l-tmp)+sign(u-tmp))));
  den=volfrac-(sum(u(setu))+sum(l(setl)))/nEl;
  lmOld=lm;
  lm=(sum(ocP(setM))/den)/nEl;
end
x=ocP(act)/lm;
[setu, setl]=deal(find(x>u),find(x<l));
x(setl)=l(setl); x(setu)=u(setu);
\end{lstlisting}
and, for the MBB beam example, this performs as shown by the green curves in \autoref{fig:optEtaandlambda}(b).

However, a closed form expression such as \eqref{eq:app-dualMap} cannot be obtained for more involved constraint expressions and therefore a root finding strategy must be employed to approximate the Lagrange multiplier. The application of \eqref{eq:app-dualMap} to the current, feasible design point ($\mathbf{x}_{(j+1)} = \mathbf{x}_{k}$) reduces to
\begin{equation}
 \label{eq:app-dualMap0}
  \lambda^{\#} = \left[ \frac{1}{mf}
  \sum^{m}_{e=1} x_{k,e}
  \left(
  -\frac{\partial_{e}c(\boldsymbol{\xi})}
  {\partial_{e}V(\boldsymbol{\xi})}
  \right)^{1/2}
  \right]^{2}
\end{equation}
since $|\mathcal{M}| = |\Omega_{h}| = m$, $|\mathcal{L}| = |\mathcal{U}| = 0$ and we made use of \eqref{eq:sensitivityComplianceVolume}. We immediately verify that \eqref{eq:app-dualMap0} is identical to \eqref{eq:estimateLM}.

\onecolumn

\section{The 2D code for compliance minimization}
 \label{Sec:2DcodeComplianceMinimization}

 \begin{lstlisting}
function top99neo(nelx,nely,volfrac,penal,rmin,ft,ftBC,eta,beta,move,maxit)
% ---------------------------- PRE. 1) MATERIAL AND CONTINUATION PARAMETERS
E0 = 1;                                                                    % Young modulus of solid
Emin = 1e-9;                                                               % Young modulus of "void"
nu = 0.3;                                                                  % Poisson ratio
penalCnt = { 1,  3, 25, 0.25 };                                            % continuation scheme on penal
betaCnt  = { 1,  2, 25,    2 };                                            % continuation scheme on beta
if ftBC == 'N', bcF = 'symmetric'; else, bcF = 0; end                      % filter BC selector
% ----------------------------------------- PRE. 2) DISCRETIZATION FEATURES
nEl = nelx * nely;                                                         % number of elements
nodeNrs = int32( reshape( 1 : (1 + nelx) * (1 + nely), 1+nely, 1+nelx ) ); % nodes numbers (defined as int32)
cVec = reshape( 2 * nodeNrs( 1 : end - 1, 1 : end - 1 ) + 1, nEl, 1 );
cMat = cVec + int32( [ 0, 1, 2 * nely + [ 2, 3, 0, 1 ], -2, -1 ] );        % connectivity matrix
nDof = ( 1 + nely ) * ( 1 + nelx ) * 2;                                    % total number of DOFs
[ sI, sII ] = deal( [ ] );
for j = 1 : 8
    sI = cat( 2, sI, j : 8 );
    sII = cat( 2, sII, repmat( j, 1, 8 - j + 1 ) );
end
[ iK , jK ] = deal( cMat( :,  sI )', cMat( :, sII )' );
Iar = sort( [ iK( : ), jK( : ) ], 2, 'descend' ); clear iK jK              % reduced assembly indexing
c1 = [12;3;-6;-3;-6;-3;0;3;12;3;0;-3;-6;-3;-6;12;-3;0;-3;-6;3;12;3;...
    -6;3;-6;12;3;-6;-3;12;3;0;12;-3;12];
c2 = [-4;3;-2;9;2;-3;4;-9;-4;-9;4;-3;2;9;-2;-4;-3;4;9;2;3;-4;-9;-2;...
    3;2;-4;3;-2;9;-4;-9;4;-4;-3;-4];
Ke = 1/(1-nu^2)/24*( c1 + nu .* c2 );                                      % lower sym. part of el. matrix
Ke0( tril( ones( 8 ) ) == 1 ) = Ke';
Ke0 = reshape( Ke0, 8, 8 );
Ke0 = Ke0 + Ke0' - diag( diag( Ke0 ) );                                    % recover full elemental matrix
% ----------------------------- PRE. 3) LOADS, SUPPORTS AND PASSIVE DOMAINS
lcDof = 2 * nodeNrs( 1, 1 );                                               % DOFs with applied load
fixed = union( 1 : 2 : 2*( nely + 1 ), 2 * nodeNrs( end, end ) );          % restrained DOFs
[ pasS, pasV ] = deal( [], [] );                                           % UD, passive solid and void el.
F = fsparse( lcDof', 1, -1, [ nDof, 1 ] );                                 % define load vector
free = setdiff( 1 : nDof, fixed );                                         % set of free DOFs
act = setdiff( (1 : nEl )', union( pasS, pasV ) );                         % set of active d.v.
% --------------------------------------- PRE. 4) DEFINE IMPLICIT FUNCTIONS
prj = @(v,eta,beta) (tanh(beta*eta)+tanh(beta*(v(:)-eta)))./...
    (tanh(beta*eta)+tanh(beta*(1-eta)));                                   % projection
deta = @(v,eta,beta) - beta * csch( beta ) .* sech( beta * ( v( : ) - eta ) ).^2 .* ...
    sinh( v( : ) * beta ) .* sinh( ( 1 - v( : ) ) * beta );                % projection eta-derivative
dprj = @(v,eta,beta) beta*(1-tanh(beta*(v-eta)).^2)./(tanh(beta*eta)+tanh(beta*(1-eta)));% proj. x-derivative
cnt = @(v,vCnt,l) v+(l>=vCnt{1})*(v<vCnt{2})*(mod(l,vCnt{3})==0)*vCnt{4};  % apply continuation
% ------------------------------------------------- PRE. 5) PREPARE FILTER
[dy,dx] = meshgrid(-ceil(rmin)+1:ceil(rmin)-1,-ceil(rmin)+1:ceil(rmin)-1);
h = max( 0, rmin - sqrt( dx.^2 + dy.^2 ) );                                % conv. kernel
Hs = imfilter( ones( nely, nelx ), h, bcF );                               % matrix of weights (filter)
dHs = Hs;
% ------------------------ PRE. 6) ALLOCATE AND INITIALIZE OTHER PARAMETERS
[ x, dsK, dV ] = deal( zeros( nEl, 1 ) );                                  % initialize vectors
dV( act, 1 ) = 1/nEl/volfrac;                                              % derivative of volume (constant)
x( act ) = ( volfrac*( nEl - length(pasV) ) - length(pasS) )/length( act );% volume fraction on active set
x( pasS ) = 1;                                                             % set x = 1 on pasS set
[ xPhys, xOld, ch, loop, U ] = deal( x, 1, 1, 0, zeros( nDof, 1 ) );       % old x, x change, it. counter, U
% ================================================= START OPTIMIZATION LOOP
while ch > 1e-6 && loop < maxit
  loop = loop + 1;                                                         % update iter. counter
  % ----------- RL. 1) COMPUTE PHYSICAL DENSITY FIELD (AND ETA IF PROJECT.)
  xTilde = imfilter( reshape( x, nely, nelx ), h, bcF ) ./ Hs;
  xPhys( act ) = xTilde( act );                                            % reshape to column vector
  if ft > 1                              % compute optimal eta* with Newton
      f = ( mean( prj( xPhys, eta, beta ) ) - volfrac ) * ( ft == 3 );     % function (volume)
      while abs( f ) > 1e-6           % Newton process for finding opt. eta
          eta = eta - f / mean( deta( xPhys( : ), eta, beta ) );
          f = mean( prj( xPhys, eta, beta ) ) - volfrac;
      end
      dHs = Hs ./ reshape( dprj( xTilde, eta, beta ), nely, nelx );        % modification of the sensitivity
      xPhys = prj( xPhys, eta, beta );                                     % projected (physical) field
  end
  ch = norm( xPhys - xOld ) ./ sqrt( nEl );
  xOld = xPhys;
  % -------------------------- RL. 2) SETUP AND SOLVE EQUILIBRIUM EQUATIONS
  sK = ( Emin + xPhys.^penal * ( E0 - Emin ) );                            % stiffness interpolation
  dsK( act ) = -penal * ( E0 - Emin ) * xPhys( act ) .^ ( penal - 1 );     % derivative of stiffness interp.
  sK = reshape( Ke( : ) * sK', length( Ke ) * nEl, 1 );
  K = fsparse( Iar( :, 1 ), Iar( :, 2 ), sK, [ nDof, nDof ] );             % assemble stiffness matrix
  U( free ) = decomposition( K( free, free ), 'chol','lower' ) \ F( free );% solve equilibrium system
  % ------------------------------------------ RL. 3) COMPUTE SENSITIVITIES
  dc = dsK .* sum( ( U( cMat ) * Ke0 ) .* U( cMat ), 2 );                  % derivative of compliance
  dc = imfilter( reshape( dc, nely, nelx ) ./ dHs, h, bcF );               % filter objective sensitivity
  dV0 = imfilter( reshape( dV, nely, nelx ) ./ dHs, h, bcF );              % filter compliance sensitivity
  % ----------------- RL. 4) UPDATE DESIGN VARIABLES AND APPLY CONTINUATION
  xT = x( act );
  [ xU, xL ] = deal( xT + move, xT - move );                               % current upper and lower bound
  ocP = xT .* real( sqrt( -dc( act ) ./ dV0( act ) ) );                    % constant part in resizing rule
  l = [ 0, mean( ocP ) / volfrac ];                                        % initial estimate for LM
  while ( l( 2 ) - l( 1 ) ) / ( l( 2 ) + l( 1 ) ) > 1e-4                   % OC resizing rule
      lmid = 0.5 * ( l( 1 ) + l( 2 ) );
      x( act ) = max( max( min( min( ocP / lmid, xU ), 1 ), xL ), 0 );
      if mean( x ) > volfrac, l( 1 ) = lmid; else, l( 2 ) = lmid; end
  end
  [penal,beta] = deal(cnt(penal,penalCnt,loop), cnt(beta,betaCnt,loop));   % apply conitnuation on parameters
  % -------------------------- RL. 5) PRINT CURRENT RESULTS AND PLOT DESIGN
  fprintf( 'It.:%5i C:%7.4f V:%7.3f ch.:%0.2e penal:%7.2f beta:%7.1f eta:%7.2f \n', ...
      loop, F'*U, mean( xPhys ), ch, penal, beta, eta ); 
  colormap( gray ); imagesc( 1 - reshape( xPhys, nely, nelx ) );
  caxis([0 1]); axis equal off; drawnow;
end
end
\end{lstlisting}


\section{The 3D code for compliance minimization}
 \label{Sec:3DcodeComplianceMinimization}
 
 \begin{lstlisting}
function top3D125(nelx,nely,nelz,volfrac,penal,rmin,ft,ftBC,eta,beta,move,maxit)
% ---------------------------- PRE. 1) MATERIAL AND CONTINUATION PARAMETERS
E0 = 1;                                                                    % Young modulus of solid
Emin = 1e-9;                                                               % Young modulus of "void"
nu = 0.3;                                                                  % Poisson ratio
penalCnt = { 1, 3, 25, 0.25 };                                             % continuation scheme on penal
betaCnt  = { 1, 2, 25,    2 };                                             % continuation scheme on beta
if ftBC == 'N', bcF = 'symmetric'; else, bcF = 0; end                      % filter BC selector
% ----------------------------------------- PRE. 2) DISCRETIZATION FEATURES
nEl = nelx * nely * nelz;                                                  % number of elements          #3D#
nodeNrs = int32( reshape( 1 : ( 1 + nelx ) * ( 1 + nely ) * ( 1 + nelz ), ...
    1 + nely, 1 + nelz, 1 + nelx ) );                                      % nodes numbering             #3D#
cVec = reshape( 3 * nodeNrs( 1 : nely, 1 : nelz, 1 : nelx ) + 1, nEl, 1 ); %                             #3D#
cMat = cVec+int32( [0,1,2,3*(nely+1)*(nelz+1)+[0,1,2,-3,-2,-1],-3,-2,-1,3*(nely+...
   1)+[0,1,2],3*(nely+1)*(nelz+2)+[0,1,2,-3,-2,-1],3*(nely+1)+[-3,-2,-1]]);% connectivity matrix         #3D#
nDof = ( 1 + nely ) * ( 1 + nelz ) * ( 1 + nelx ) * 3;                     % total number of DOFs        #3D#
[ sI, sII ] = deal( [ ] );
for j = 1 : 24
    sI = cat( 2, sI, j : 24 );
    sII = cat( 2, sII, repmat( j, 1, 24 - j + 1 ) );
end
[ iK , jK ] = deal( cMat( :,  sI )', cMat( :, sII )' );
Iar = sort( [ iK( : ), jK( : ) ], 2, 'descend' ); clear iK jK              % reduced assembly indexing
Ke = 1/(1+nu)/(2*nu-1)/144 *( [ -32;-6;-6;8;6;6;10;6;3;-4;-6;-3;-4;-3;-6;10;...
    3;6;8;3;3;4;-3;-3; -32;-6;-6;-4;-3;6;10;3;6;8;6;-3;-4;-6;-3;4;-3;3;8;3;...
    3;10;6;-32;-6;-3;-4;-3;-3;4;-3;-6;-4;6;6;8;6;3;10;3;3;8;3;6;10;-32;6;6;...
    -4;6;3;10;-6;-3;10;-3;-6;-4;3;6;4;3;3;8;-3;-3;-32;-6;-6;8;6;-6;10;3;3;4;...
    -3;3;-4;-6;-3;10;6;-3;8;3;-32;3;-6;-4;3;-3;4;-6;3;10;-6;6;8;-3;6;10;-3;...
    3;8;-32;-6;6;8;6;-6;8;3;-3;4;-3;3;-4;-3;6;10;3;-6;-32;6;-6;-4;3;3;8;-3;...
    3;10;-6;-3;-4;6;-3;4;3;-32;6;3;-4;-3;-3;8;-3;-6;10;-6;-6;8;-6;-3;10;-32;...
    6;-6;4;3;-3;8;-3;3;10;-3;6;-4;3;-6;-32;6;-3;10;-6;-3;8;-3;3;4;3;3;-4;6;...
    -32;3;-6;10;3;-3;8;6;-3;10;6;-6;8;-32;-6;6;8;6;-6;10;6;-3;-4;-6;3;-32;6;...
    -6;-4;3;6;10;-3;6;8;-6;-32;6;3;-4;3;3;4;3;6;-4;-32;6;-6;-4;6;-3;10;-6;3;...
    -32;6;-6;8;-6;-6;10;-3;-32;-3;6;-4;-3;3;4;-32;-6;-6;8;6;6;-32;-6;-6;-4;...
    -3;-32;-6;-3;-4;-32;6;6;-32;-6;-32]+nu*[ 48;0;0;0;-24;-24;-12;0;-12;0;...
    24;0;0;0;24;-12;-12;0;-12;0;0;-12;12;12;48;0;24;0;0;0;-12;-12;-24;0;-24;...
    0;0;24;12;-12;12;0;-12;0;-12;-12;0;48;24;0;0;12;12;-12;0;24;0;-24;-24;0;...
    0;-12;-12;0;0;-12;-12;0;-12;48;0;0;0;-24;0;-12;0;12;-12;12;0;0;0;-24;...
    -12;-12;-12;-12;0;0;48;0;24;0;-24;0;-12;-12;-12;-12;12;0;0;24;12;-12;0;...
    0;-12;0;48;0;24;0;-12;12;-12;0;-12;-12;24;-24;0;12;0;-12;0;0;-12;48;0;0;...
    0;-24;24;-12;0;0;-12;12;-12;0;0;-24;-12;-12;0;48;0;24;0;0;0;-12;0;-12;...
    -12;0;0;0;-24;12;-12;-12;48;-24;0;0;0;0;-12;12;0;-12;24;24;0;0;12;-12;...
    48;0;0;-12;-12;12;-12;0;0;-12;12;0;0;0;24;48;0;12;-12;0;0;-12;0;-12;-12;...
    -12;0;0;-24;48;-12;0;-12;0;0;-12;0;12;-12;-24;24;0;48;0;0;0;-24;24;-12;...
    0;12;0;24;0;48;0;24;0;0;0;-12;12;-24;0;24;48;-24;0;0;-12;-12;-12;0;-24;...
    0;48;0;0;0;-24;0;-12;0;-12;48;0;24;0;24;0;-12;12;48;0;-24;0;12;-12;-12;...
    48;0;0;0;-24;-24;48;0;24;0;0;48;24;0;0;48;0;0;48;0;48 ] );             % elemental stiffness matrix  #3D#
Ke0( tril( ones( 24 ) ) == 1 ) = Ke';
Ke0 = reshape( Ke0, 24, 24 );
Ke0 = Ke0 + Ke0' - diag( diag( Ke0 ) );                                    % recover full matrix
% ----------------------------- PRE. 3) LOADS, SUPPORTS AND PASSIVE DOMAINS
lcDof = 3 * nodeNrs( 1 : nely + 1, 1, nelx + 1 );
fixed = 1 : 3 * ( nely + 1 ) * ( nelz + 1 );
[ pasS, pasV ] = deal( [], [] );                                           % passive solid and void elements
F = fsparse( lcDof, 1, -sin((0:nely)/nely*pi)', [ nDof, 1 ] );             % define load vector
free = setdiff( 1 : nDof, fixed );                                         % set of free DOFs
act = setdiff( ( 1 : nEl )', union( pasS, pasV ) );                        % set of active d.v.
% --------------------------------------- PRE. 4) DEFINE IMPLICIT FUNCTIONS
prj = @(v,eta,beta) (tanh(beta*eta)+tanh(beta*(v(:)-eta)))./...
    (tanh(beta*eta)+tanh(beta*(1-eta)));                                   % projection
deta = @(v,eta,beta) - beta * csch( beta ) .* sech( beta * ( v( : ) - eta ) ).^2 .* ...
    sinh( v( : ) * beta ) .* sinh( ( 1 - v( : ) ) * beta );                % projection eta-derivative 
dprj = @(v,eta,beta) beta*(1-tanh(beta*(v-eta)).^2)./(tanh(beta*eta)+tanh(beta*(1-eta)));% proj. x-derivative
cnt = @(v,vCnt,l) v+(l>=vCnt{1}).*(v<vCnt{2}).*(mod(l,vCnt{3})==0).*vCnt{4};
% -------------------------------------------------- PRE. 5) PREPARE FILTER
[dy,dz,dx]=meshgrid(-ceil(rmin)+1:ceil(rmin)-1,...
    -ceil(rmin)+1:ceil(rmin)-1,-ceil(rmin)+1:ceil(rmin)-1 );
h = max( 0, rmin - sqrt( dx.^2 + dy.^2 + dz.^2 ) );                        % conv. kernel                #3D#
Hs = imfilter( ones( nely, nelz, nelx ), h, bcF );                         % matrix of weights (filter)  #3D#
dHs = Hs;
% ------------------------ PRE. 6) ALLOCATE AND INITIALIZE OTHER PARAMETERS
[ x, dsK, dV ] = deal( zeros( nEl, 1 ) );                                  % initialize vectors
dV( act, 1 ) = 1/nEl/volfrac;                                              % derivative of volume
x( act ) = ( volfrac*( nEl - length(pasV) ) - length(pasS) )/length( act );% volume fraction on active set
x( pasS ) = 1;                                                             % set x = 1 on pasS set
[ xPhys, xOld, ch, loop, U ] = deal( x, 1, 1, 0, zeros( nDof, 1 ) );       % old x, x change, it. counter, U
% ================================================= START OPTIMIZATION LOOP
while ch > 1e-6 && loop < maxit
  loop = loop + 1;                                                         % update iter. counter
  % ----------- RL. 1) COMPUTE PHYSICAL DENSITY FIELD (AND ETA IF PROJECT.)
  xTilde = imfilter( reshape( x, nely, nelz, nelx ) ./ Hs, h, bcF );       % filtered field              #3D#
  xPhys( act ) = xTilde( act );                                            % reshape to column vector
  if ft > 1                              % compute optimal eta* with Newton
      f = ( mean( prj( xPhys, eta, beta ) ) - volfrac )  * (ft == 3);      % function (volume)
      while abs( f ) > 1e-6           % Newton process for finding opt. eta
          eta = eta - f / mean( deta( xPhys, eta, beta ) );
          f = mean( prj( xPhys, eta, beta ) ) - volfrac;
      end
      dHs = Hs ./ reshape( dprj( xPhys, eta, beta ), nely, nelz, nelx );   % sensitivity modification    #3D#
      xPhys = prj( xPhys, eta, beta );                                     % projected (physical) field
  end
  ch = norm( xPhys - xOld ) ./ nEl;
  xOld = xPhys;
  % -------------------------- RL. 2) SETUP AND SOLVE EQUILIBRIUM EQUATIONS
  sK = ( Emin + xPhys.^penal * ( E0 - Emin ) );
  dsK( act ) = -penal * ( E0 - Emin ) * xPhys( act ) .^ ( penal - 1 );
  sK = reshape( Ke( : ) * sK', length( Ke ) * nEl, 1 );
  K = fsparse( Iar( :, 1 ), Iar( :, 2 ), sK, [ nDof, nDof ] );
  L = chol( K( free, free ), 'lower' );
  U( free ) = L' \ ( L \ F( free ) );                                      % f/b substitution
  % ------------------------------------------ RL. 3) COMPUTE SENSITIVITIES
  dc = dsK .* sum( ( U( cMat ) * Ke0 ) .* U( cMat ), 2 );                  % derivative of compliance
  dc = imfilter( reshape( dc, nely, nelz, nelx ), h, bcF ) ./ dHs;         % filter objective sens.      #3D#
  dV0 = imfilter( reshape( dV, nely, nelz, nelx ), h, bcF ) ./ dHs;        % filter compliance sens.     #3D#
  % ----------------- RL. 4) UPDATE DESIGN VARIABLES AND APPLY CONTINUATION
  xT = x( act );
  [ xU, xL ] = deal( xT + move, xT - move );                               % current upper and lower bound
  ocP = xT( act ) .* sqrt( - dc( act ) ./ dV0( act ) );                    % constant part in resizing rule
  l = [ 0, mean( ocP ) / volfrac ];                                        % initial estimate for LM
  while ( l( 2 ) - l( 1 ) ) / ( l( 2 ) + l( 1 ) ) > 1e-4                   % OC resizing rule
      lmid = 0.5 * ( l( 1 ) + l( 2 ) );
      x( act ) = max( max( min( min( ocP/lmid, xU ), 1 ), xL ), 0 );
      if mean( x ) > volfrac, l( 1 ) = lmid; else, l( 2 ) = lmid; end
  end
  [penal,beta] = deal(cnt(penal,penalCnt,loop), cnt(beta,betaCnt,loop));   % apply conitnuation on parameters
  % -------------------------- RL. 5) PRINT CURRENT RESULTS AND PLOT DESIGN
  fprintf( 'It.:%5i C:%6.5e V:%7.3f ch.:%0.2e penal:%7.2f beta:%7.1f eta:%7.2f lm:%0.2e \n', ...
      loop, F'*U, mean(xPhys(:)), ch, penal, beta, eta, lmid );
  isovals = shiftdim( reshape( xPhys, nely, nelz, nelx ), 2 );
  isovals = smooth3( isovals, 'box', 1 );
  patch(isosurface(isovals, .5),'FaceColor','b','EdgeColor','none');
  patch(isocaps(isovals, .5),'FaceColor','r','EdgeColor','none');
  drawnow; view( [ 145, 25 ] ); axis equal tight off; cla();
end
end
\end{lstlisting}

%
\end{document}